\documentclass[journal]{IEEEtran}
\ifCLASSINFOpdf
   \usepackage[pdftex]{graphicx}
   \graphicspath{{Figures/}}
\else
\fi

\usepackage{glossaries}
\newacronym{ai}{AI}{All Intra}
\newacronym{hevc}{HEVC}{High Efficiency Video Coding}
\newacronym{avc}{AVC}{Advanced Video Coding}
\newacronym{jvet}{JVET}{Joint Video Experts Team}
\newacronym{cfp}{CfP}{Call for Proposal}

\newacronym{mpeg}{MPEG}{Motion Picture Experts Group}
\newacronym{vceg}{VQEG}{Video Coding Experts Group}

\newacronym{asic}{ASIC}{Application-Specific Integrated Circuit}
\newacronym{fpga}{FPGA}{Field-Programmable Gate Array}
\newacronym{jem}{JEM}{Joint Exploration Model}
\newacronym{mts}{MTS}{Multiple Transform Selection}
\newacronym{lfnst}{LFNST}{Low Frequency Non-Separable Transform}
\newacronym{dct}{DCT}{Discrete Cosine Transform}
\newacronym{dst}{DST}{Discrete Sine Transform}
\newacronym{ctc}{CTC}{Common Test Conditions}
\newacronym{fft}{FFT}{Fast Fourier Transform}

\newacronym{sps}{SPS}{Sequence Parameter Set}
\newacronym{dsp}{DSP}{Digital Signal Processing}
\newacronym{dft}{DFT}{Discrete Fourier Transform}
\newacronym{ip}{IP}{Intellectual Property}
  
\newacronym{ra}{RA}{Random Access}
\newacronym{vvc}{VVC}{Versatile Video Coding}
\newacronym{vtm}{VTM}{VVC Test Model}
\newacronym{iict}{IICT}{Inverse Integer Core Transforms}
\newacronym{vlsi}{VLSI}{Very Large Scale Integration}
\newacronym{bdr}{BD-BR}{Bj\o ntegaard Delta Rate}
\newacronym{sram}{SRAM}{Static Random-Access Memory}

\newacronym{ram}{RAM}{Random-Access Memory}
\newacronym{rom}{ROM}{Read-Only Memory}
\newacronym{rm}{RM}{Regular Multiplier}
\newacronym{mcm}{MCM}{Multiple Constant Multiplier}

\newacronym{dc}{DC}{Design Compiler}
\newacronym{tsmc}{TSMC}{Taiwan Semiconductor Manufacturing Company}
\newacronym{idst}{IDST}{Inverse DST}
\newacronym{idct}{IDCT}{Inverse DCT}

\newacronym{ctu}{CTU}{Coding Tree Unit}
\newacronym{ctb}{CTB}{Coding Tree Block}
\newacronym{alm}{ALM}{Adaptive Logic Module}

\usepackage{bbding} 
\usepackage{multirow}
\newcommand{\Figure}[1] {Fig.~#1}

 \newcommand{\plh}{%
  {{\mkern-1mu\times\mkern-1.5mu}}%
}
\usepackage{amsmath,amssymb,amsfonts}
\usepackage{xcolor,colortbl}
\usepackage{makecell, soul}
\definecolor{Gray}{gray}{1}
\usepackage{hyperref}       
\usepackage{pifont}
\usepackage{cite}

\usepackage{xcolor,colortbl}
\definecolor{Gray}{gray}{0.90}
\definecolor{Gray1}{gray}{0.95}
\newcommand{\adcomment}[1]{{\color{black}{#1}}}
\definecolor{intra}{rgb}{0.84, 0.92, 0.91} 
\begin{document}
%

\title{Lightweight Hardware Transform Design for the Versatile Video Coding 4K ASIC Decoders}
%
%
%

\author{Ibrahim Farhat,
        Wassim Hamidouche,
        Adrien Grill,
        Daniel Ménard,
        and~Olivier Déforges
\thanks{Ibrahim~Farhat, Wassim Hamidouche, Daniel Menard and Olivier Déforges are with Univ. Rennes, INSA Rennes, CNRS, IETR - UMR 6164, 20 Avenue des Buttes de Coesmes, 35708 Rennes, France. E-mails: \href{mailto:Ibrahim.Farhat@insa-rennes.fr}{Ibrahim.Farhat@insa-rennes.fr}, \href{mailto:wassim.hamidouche@insa-rennes.fr}{wassim.hamidouche@insa-rennes.fr}, \href{mailto:Daniel.Menard@insa-rennes.fr}{Daniel.Menard@insa-rennes.fr},  \href{mailto:Olivier.Deforges@insa-rennes.fr}{Olivier.Deforges@insa-rennes.fr}.}
\thanks{Ibrahim~Farhat and Adrien~Grill are with  VITEC, 99 rue Pierre Semard, 92320 Chatillon, France. E-mails: \href{mailto:ibrahim.farhat@vitec.com}{ibrahim.farhat@vitec.com}, \href{mailto:adrien.grill@vitec.com}{adrien.grill@vitec.com}}
}

%
%

\markboth{IEEE Transactions on Consumer Electronics, 2021}%
{Shell \MakeLowercase{\textit{et al.}}: Bare Demo of IEEEtran.cls for IEEE Journals}
%



\maketitle

\begin{abstract}
\Gls{vvc} is the next generation video coding standard finalized in July 2020. \Gls{vvc} introduces new coding tools enhancing the coding efficiency compared to its predecessor \gls{hevc}. These new tools have a significant impact on the \gls{vvc} software decoder complexity estimated to 2 times HEVC decoder complexity. In particular, the transform module includes in \gls{vvc} separable and non-separable transforms named \gls{mts} and \gls{lfnst} tools, respectively.        
In this paper, we present an area-efficient hardware architecture of the inverse transform module for a \gls{vvc} decoder. The proposed design uses a total of 64 regular multipliers in a pipelined architecture targeting \gls{asic} platforms. It consists in a multi-standard architecture that supports the transform modules of recent MPEG standards including \gls{avc},  \gls{hevc} and \gls{vvc}. The architecture leverages all primary and secondary transforms' optimisations including butterfly decomposition, coefficients zeroing and the inherent linear relationship between the transforms. The synthesized results show that the proposed method sustains a constant throughput of 1 sample per cycle and a constant latency for all block sizes. The proposed hardware inverse transform module operates at 600 MHz frequency enabling to decode in real-time 4K video at 30 frames per second in 4:2:2 chroma sub-sampling format. The proposed module has been integrated in an \gls{asic} UHD decoder targeting energy-aware decoding of \gls{vvc} videos on consumer devices.  
\end{abstract}


%
\IEEEpeerreviewmaketitle

\glsresetall

\section{Introduction}
\IEEEPARstart{T}{he} \gls{vvc} is the next generation video coding standard jointly developed by \gls{mpeg} and \gls{vceg} under the \gls{jvet}. \gls{vvc} was finalized in July 2020 as ITU-T H.266 | MPEG-I - Part 3 (ISO/IEC 23090-3) standard~\cite{9328514, hamidouche2021versatile}. It introduces several new coding tools enabling up to 40\%  of coding gains beyond the \gls{hevc} standard~\cite{JVET-O0003, PCS-2019-VVC-QE}. The \gls{vvc} transform module includes two new tools called \gls{mts} and \gls{lfnst}~\cite{9449858}. The \gls{mts} tool involves three trigonometrical transform types including \gls{dct} type II (DCT-II), \gls{dct} type VIII (DCT-VIII) and \gls{dst} type VII (DST-VII) with block sizes that can reach $64 \plh 64$ for DCT-II and $32 \plh 32$ for DCT-VIII and DST-VII. The use of \gls{dct}/\gls{dst} families gives the ability to apply separable transforms, the transformation of a block can be applied separately in horizontal and vertical directions. Therefore, the \gls{vvc} encoder selects a combination of horizontal and vertical transforms that minimizes the rate-distortion cost $J$, computed in~(\ref{eq1:rdo}) as a trade-off between distortion $D$ and rate $R$
\begin{equation}
    J = D + \lambda \, R,
    \label{eq1:rdo}
\end{equation}
where $\lambda$ is a Lagrangian parameter computed depending on the quantisation parameter.  

\begin{figure}[h]
\includegraphics[width=0.48\textwidth]{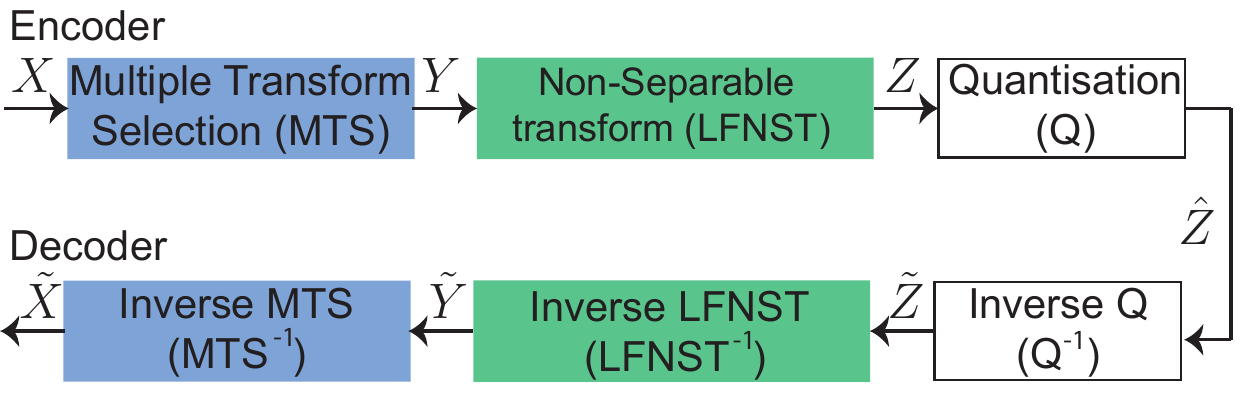}
\caption{VVC transform module}
\label{fig:vvc-trasform}
\end{figure}

The \gls{lfnst} is applied after the separable transform and before the quantisation at the encoder side. At the decoder side, it is applied after the inverse quantisation and before the inverse separable transform. \Figure{\ref{fig:vvc-trasform}} illustrates the \gls{vvc} transform module at both encoder and decoder sides. The transform module relies on matrix multiplication with $\mathcal{O} (N^3)$ and $ \mathcal{O}(N^4)$ computing complexities for separable and non-separable transforms, respectively. In terms of memory usage, the \gls{vvc} transform module requires higher memory allocated to store the coefficients of the transform kernels: three kernels are defined for \gls{mts} and eight kernels for \gls{lfnst}.

Fast computing algorithms for \glspl{dct}/\glspl{dst} have been widely investigated in the literature. The main objective of these algorithms is to perform the transform with the lowest number of multiplications compared to a naive matrix multiplication requiring for $N \plh N$ square matrix $N^3$ multiplications (ie. $\mathcal{O} (N^3)$ computational complexity). The \gls{dct}-II can be decomposed in butterfly~\cite{PLONKA2005309, 266596, VETTERLI1984267} reducing computational complexity in terms of number of multiplications and additions. This decomposition is hardware-friendly enabling hardware resources sharing between blocks of different sizes for both computation and memory usage. In contrast to \gls{dct}-II, \gls{dst}-VII/\gls{dct}-VIII have less efficient fast implementation algorithms~\cite{doi:10.1117/12.903685, 4471889, DST-7-DCC-2019} and dot not enable hardware resources sharing. 
Table~\ref{table::complex} summarises the performance in terms of number of multiplications and additions of fast implementations of \gls{dct}-II and \gls{dst}-VIII at different sizes $N \in \{8, \, 16, \, 32, \, 64 \}$.  

\begin{table*}[t]
\renewcommand{\arraystretch}{1.3}
\caption{Complexity performance of the \gls{dct}-II and \gls{dst}-VII fast computing algorithms}
\label{table::complex}  
\centering
\begin{tabular}{l | c c c c | c c c c | c c c c | c c c c}
\hline
\hline
\multirow{2}{*}{Transforms} 
& \multicolumn{4}{c}{$N=8$} & \multicolumn{4}{c}{$N=16$} & \multicolumn{4}{c}{$N=32$} & \multicolumn{4}{c}{$N=64$} \\ 
 \cline{2-17}
 &Ref. & + & $\times$ & All & Ref. & + & $\times$  & All &  Ref. & + & $\times$  & All &  Ref. & + & $\times$  & All   \\ 
\hline
 \gls{dct}-II &  \cite{266596} & $29$ & $11$ & $40$ &  \cite{266596} & $81$ & $31$ & $112$ & \cite{VETTERLI1984267} & $209$ & $80$ & $289$ &  \cite{VETTERLI1984267} & $192$ & $513$ & $707$ \\ 
 \gls{dct}-II (HEVC) & \cite{6575105}  & $37$ & $24$ & $61$ & \cite{6575105} & $113$ & $86$ & $199$ & \cite{6575105} & $401$ & $342$ & $743$ & \cite{6575105} & $807$ & $683$ & $1490$ \\ 
 \gls{dst}-VII & \cite{doi:10.1117/12.903685} &$77$ & $21$  & $98$ & \cite{4471889}& $150$ & $146$ & $296$ & $-$ & $-$ & $-$ & & $-$  & $-$  & $-$ & \\ 
 \gls{dst}-VII & \cite{DST-7-DCC-2019} &$-$ & $-$ & & \cite{DST-7-DCC-2019}& $155$ & $127$ & $282$ &  \cite{DST-7-DCC-2019} & $718$ & $620$ & $1338$ & \cite{DST-7-DCC-2019}  & $2331$  & $2207$ & $4538$ \\ 
   \gls{dst}-VII & \cite{8649728} &$77$ & $21$  & $98$ & \cite{8649728} &  $125$ & $42$ & $167$ & \cite{8649728}  & $279$ & $93$ & $372$ & $-$  &  $-$  &  $-$ & \\ 
 Matrix Multip.  & $-$  & $56$& $64$ & $120$ & $-$ & $240$& $256$ & $496$ & $-$ & $992$& $1024$& $2016$ & $-$ & $4032$ & $4096$  & $8128$ \\ 
\hline
\hline
\end{tabular}
\end{table*}

This paper addresses a hardware implementation of the \gls{vvc} inverse transform module for \gls{asic} platforms. To the best of our knowledge, this is the first hardware design that includes both separable and non-separable transforms supporting all block sizes. The proposed implementation relies on a shared multipliers architecture using 32 multipliers for the separable transform and the same for the non-separable one. It also exploits all recursion and decomposition properties of the considered transforms to optimize the hardware resources and speedup the design. The choice of the shared multipliers architecture was based on our previous study recently published by Farhat et al.~\cite{9054281}. This latter compares between two different hardware architectures for the \gls{mts}, one uses Regular Multipliers (RM) while the second uses Multiple Constant Multipliers (MCM). This study showed better performance for the RM based architecture compared to MCM one for this particular problem. The main objective of this paper is to propose an optimized hardware implementation of the \gls{vvc} inverse transform blocks for \gls{asic} decoder featuring many advantages: 

\begin{itemize}
\item Minimise the hardware area by leveraging all possible optimisations offered by the transforms.
\item Support all block sizes of both \gls{mts} and \gls{lfnst}. 
\item Support all recent \gls{mpeg} standards including AVC, \gls{hevc} and \gls{vvc}.  
\item Scalable design enabling to enhance both throughput and latency by only increasing the considered number of regular multipliers.    
\end{itemize}

The proposed \gls{vvc} inverse transform hardware design is able to sustain a constant latency and throughput regardless the coding configuration and the tools selected by the encoder. We target a fixed throughput of 1 sample/cycle for the entire transformation process with 2 samples/cycle for the 1D separable transform (\gls{mts}) and a 2 samples/cycle for the non-separable transform (\gls{lfnst}) while sustaining a fixed system latency for all transform sizes and types. This latter constraint is importance to accurately predict the performance of the process and facilitate chaining between transform blocks. The proposed inverse transform design has been successfully integrated into a hardware multi-standard \gls{asic} decoder. This latter can be embedded on various consumer electronics devices such as mobiles phones, setup boxes, TVs, etc.

The rest of this paper is organized as follows. Section~\ref{sec:background} presents the background on the \gls{vvc} transform module including \gls{mts} and \gls{lfnst}.  The existing hardware implementations of the transforms module are presented in Section~\ref{sec:related-works}. The proposed hardware implementations of \gls{mts} and \gls{lfnst} blocks are investigated in Section~\ref{sec:proposed-solution}. In Section~\ref{sec:results}, the performance of the proposed hardware module is assessed in terms of speedup and hardware area. Finally, Section~\ref{sec:conclusion} concludes the paper.

\section{Background}
\label{sec:background}
In this section we describe the \gls{vvc} transform module including \gls{mts} and \gls{lfnst} blocks.

\subsection{Separable transform module}
The concept of separable 2D transform enables applying two 1D transforms separately in horizontal and vertical directions
\begin{equation}
Y = T_V \cdot X \cdot T_H^T,  
\end{equation}
where $X$ is the input residual matrix of size $M \plh N$, $T_H$ and $T_V$ are horizontal and vertical transform matrices of sizes $N \plh N$ and $M \plh M$, respectively. "$\cdot$" stands for matrix multiplication. 

The inverse 2D separable transform is expressed in~(\ref{equ:inv2Dtran}). 
\begin{equation} 
\tilde{X} =  T_V^T \cdot \tilde{Y} \cdot T_H.
\label{equ:inv2Dtran}
\end{equation} 

The \gls{hevc} transform module involves the \gls{dct}-II for square blocks of size $N \plh N$ with $N \in \{ 8, \, 16, \, 32 \}$ and \gls{dst}-VII for square block of size $4\plh4$~\cite{6575105, 6522806}. The concept of transform competition allows testing different transforms and types to select the one that minimises the rate distortion cost. It enables adapting the transformation to the statistics of the encoded signal (ie. block of residuals). The transform competition has been investigated under \gls{hevc} standard~\cite{7471919, 8291719}, and more recently under the \gls{jem} software~\cite{8281012}.  The \gls{jem} codec defines five trigonometrical transform types including \gls{dct}-II, V and VIII, and \gls{dst}-I and VII. Adopting transform competition under the \gls{jem} brings significant coding efficiency enhancements estimated between $3\%$ and $5\%$ of bitrate reductions~\cite{8281012}. This coding gain is achieved at the expense of complexity overhead required to test the transform candidates at the encoder side. Moreover, additional memory is required at both encoder and decoder to store the coefficients of those transform kernels. To cope with the complexity issue, subsets of transform candidates are defined offline, and only a subset of transforms are tested at the encoder depending on the block partitioning and prediction configuration such as the block size and Intra prediction mode~\cite{8281012}, respectively.

The \gls{mts} block in \gls{vvc} involves three transform types including \gls{dct}-II, VIII and \gls{dst}-VII.  The kernels of DCT-II $C_2$, DST-VII $S_7$ and DCT-VIII $C_8$ are derived from ~(\ref{Equ:dct-2}), (\ref{Equ:dst-7}) and (\ref{Equ:dct-8}), respectively.  
\begin{equation}
C^N_{2\, i, j} = \gamma_i \sqrt{\frac{2}{N}} \cos \left (  \frac{\pi (i-1)(2j-1)}{2N}  \right ) ,
\label{Equ:dct-2}
\end{equation}
with $\gamma_i=\left\{ \begin{array}{cc}
\sqrt{\frac{1}{2}} & i=1, \\ 
1 & i\in \{2, \dots ,N \}. \end{array}
\right.$. 

\begin{equation}
S^N_{7\, i,j}  = \sqrt{\frac{4}{2N+1}} \sin \left ( \frac{\pi (2i-1)j}{2N+1} \right).
\label{Equ:dst-7}
\end{equation}

\begin{equation}
C^N_{8\, i, j} = \sqrt{\frac{4}{2N+1}} \cos \left (  \frac{\pi (2i-1) (2j-1)}{2(2N+1)} \right ), 
\label{Equ:dct-8}
\end{equation} 
with $(i, j) \in \{1, 2, \dots, N \}^2$ and $N$ is the transform size.

\begin{figure}[t]
\includegraphics[width=0.48\textwidth]{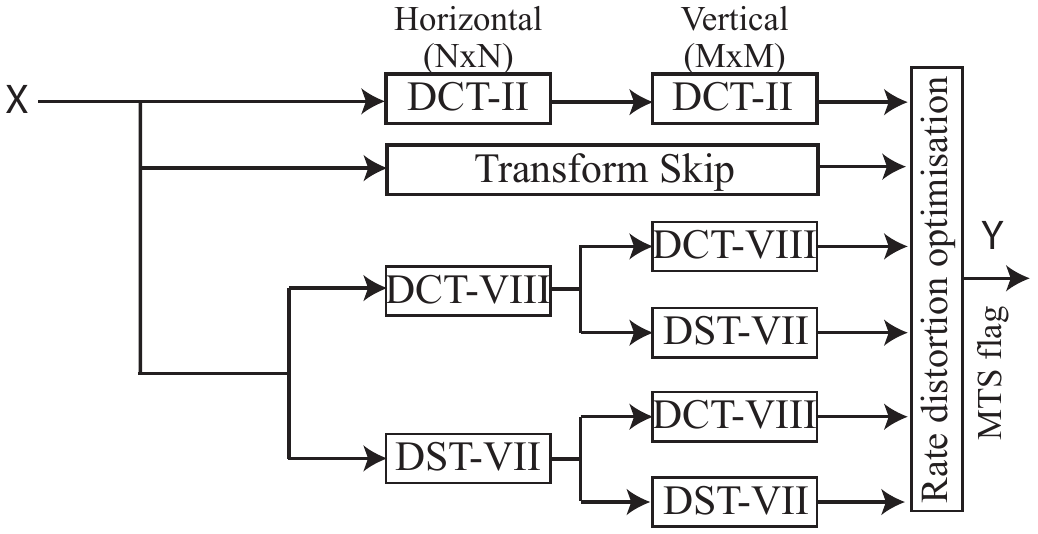}
\caption{The concept of 2D separable  transforms selection in VVC. $X$ is the input block of residuals, $Y$ is the output transformed block and MTS flag is the index of the selected set of transforms. \adcomment{The DST-VII and DCT-VIII transforms are used only for Luma samples of block size lower than 64.}}
\label{fig:vvc-MTS-module}
\end{figure}

As illustrated in \Figure{\ref{fig:vvc-MTS-module}}, the \gls{mts} concept selects, for Luma blocks of size lower than 64, a set of transforms that minimizes the rate distortion cost among five transform sets and the skip configuration. However, only \gls{dct}-II is considered for chroma components and Luma blocks of size 64. The { \it sps\_mts\_enabled\_flag} flag defined at the \gls{sps} enables to activate the \gls{mts} concept at the encoder side. Two other flags are defined at the \gls{sps} level to signal whether implicit or explicit \gls{mts} signalling is used for Intra and Inter coded blocks, respectively. For the explicit signalling, used by default in the reference software, the {\it tu\_mts\_idx } syntax element signals the selected horizontal and vertical transforms as specified in Table~\ref{table::MTS_synt}. This flag is coded with Truncated Rice code with rice parameter $p=0$ and $cMax =4$ (TRp). 
To reduce the computational cost of large block–size transforms, the effective height $M^\prime$ and width $N^\prime$ of the coding block (CB) are reduced depending of the CB size and transform type
\begin{equation}
N^\prime =\left\{ \begin{array}{cc}
min(N, 16) & trTypeHor > 0, \\ 
min(N, 32) & \text{otherwise}. \end{array}
\right.
\label{Equ:NNzeroH}
\end{equation}

\begin{equation}
M^\prime =\left\{ \begin{array}{cc}
min(M, 16) & trTypeVer > 0, \\ 
min(M, 32) & \text{otherwise}. \end{array}
\right.
\label{Equ:NNzeroV}
\end{equation}

\begin{figure}[t]
\includegraphics[width=0.48\textwidth]{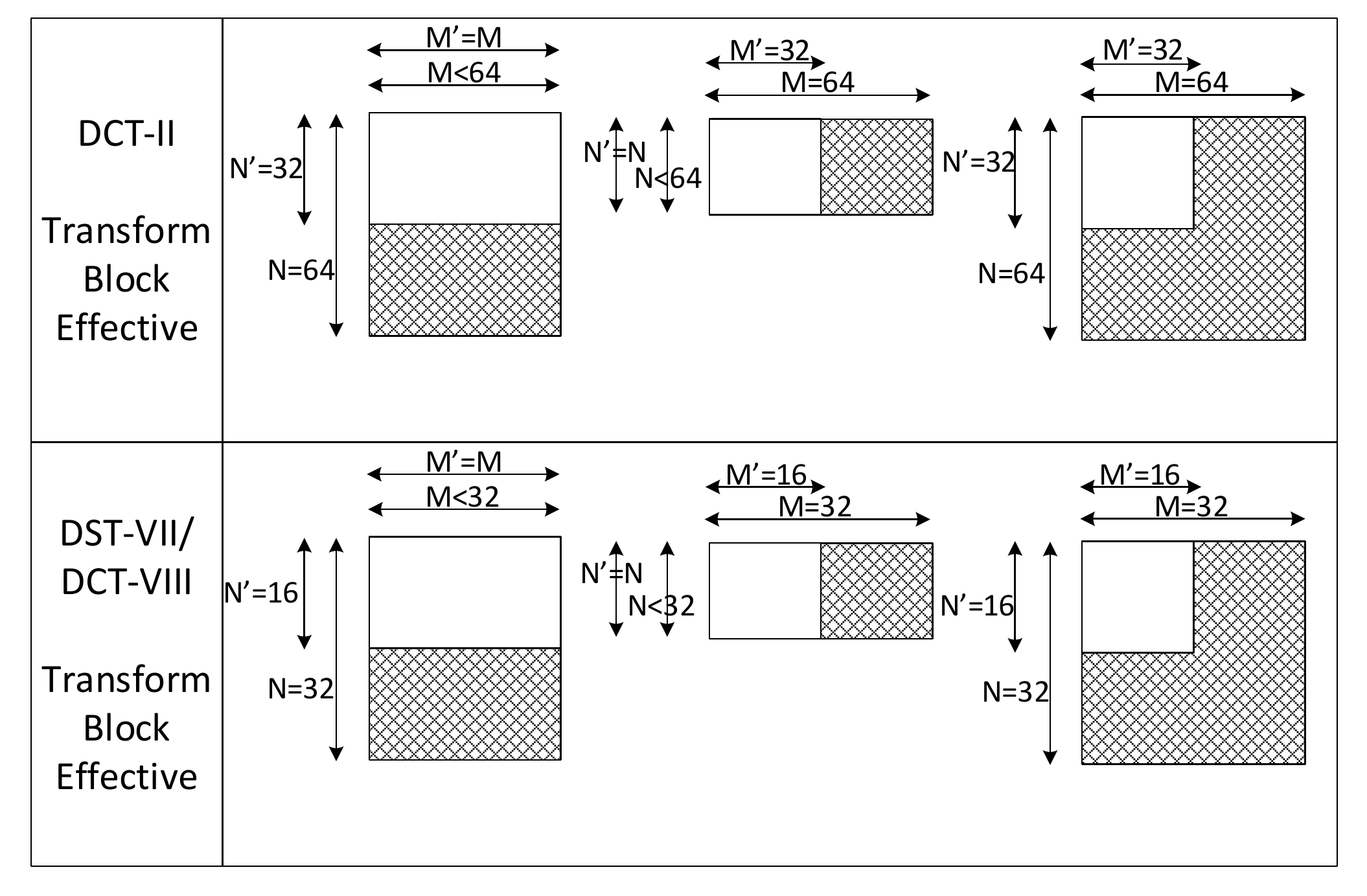}
\caption{Zeroing for large block sizes, $M^\prime$ and $N^\prime$ are the effective width and height sizes. \adcomment{The block area set to zero is illustrated in gray color}. }
\label{fig:vvc-zeroing}
\end{figure}

In~(\ref{Equ:NNzeroH}) and (\ref{Equ:NNzeroV}), $M^\prime$ and $N^\prime$ are the effective width and height sizes, $trTypeHor$ and $trTypeVer$ are respectively the types of vertical and horizontal transforms (0: \gls{dct}-II, 1: \gls{dct}-VIII and 2: \gls{dst}-VII), and the $min(a,b)$ function returns the minimum between $a$ and $b$. The sample value beyond the limits of the effective $N$ and $M$ are considered to be zero, thus reducing the computational cost of the 64-size \gls{dct}-II and 32-size \gls{dct}-VIII/\gls{dst}-VII transforms. This concept is called zeroing in the \gls{vvc} standard. \Figure{\ref{fig:vvc-zeroing}} shows the possible zeroing scenarios for \gls{dct}-II and \gls{dct}-VIII/\gls{dst}-VII transforms.

\begin{table}
\centering
\caption{Primary Transform signaling in VVC}
\label{table::MTS_synt}  
{ \renewcommand{\arraystretch}{1.2}%
\begin{tabular}{c | c | c}
\hline
   \hline
\multirow{2}{*}{\it tu\_mts\_idx}   &  \multicolumn{2}{c}{Transform Direction}  \\
 \cline{2-3}
    &  Horizontal Transform & Vertical Transform  \\
 
    \hline
    0 & \gls{dct}-II & \gls{dct}-II \\ 
    1 & \gls{dst}-VII & \gls{dst}-VII \\ 
    2 & \gls{dct}-VIII & \gls{dst}-VII \\ 
    3 & \gls{dst}-VII & \gls{dct}-VIII \\ 
    4 & \gls{dct}-VIII & \gls{dct}-VIII \\ 
\hline
   \hline
\end{tabular}}
\end{table}
\subsection{Non-separable transform module}
\label{sec:lfnst}
The Low-Frequency Non-Separable Transform (LFNST)~\cite{7906344, LFNST-PCS} has been adopted in the VTM version 5.  The \gls{lfnst} relies on matrix multiplication applied between the forward primary transform and the quantisation at the encoder side:
\begin{equation}
\vec{Z} = T \cdot \vec{Y},
\end{equation}
where the vector $\vec{Y}$ includes the coefficients of the residual block rearranged in a vector and the matrix $T$ contains the coefficients transform kernel. The \gls{lfnst} is enabled only when \gls{dct}-II is used as a primary transform. 
The inverse \gls{lfnst} is expressed in~(\ref{equ:invsep-trans}).
\begin{equation}
\vec{\tilde{Y}} = T^T \cdot \vec{\tilde{Z}}.
\label{equ:invsep-trans}
\end{equation}

Four sets of two \gls{lfnst} kernels of sizes $16\plh16$ and $64\plh64$ are applied on 16 coefficients of small blocks (min (width, height) $<$ 8 ) and 64 coefficients of larger blocks (min (width, height) $>$ 4), respectively. The \gls{vvc} specification defines four different transform sets selected depending on the Intra prediction mode and each set defines two transform kernels. The used kernel within a set is signalled in the bitstream. Table~\ref{table::LFNST_index} gives the transform set index depending on the Intra prediction mode. The transform index within a set is coded with a Truncated Rice code with rice parameter $p=0$ and $cMax =2$ (TRp) and only the first bin is context coded. The \gls{lfnst} is applied on Intra CU for both Intra and Inter slices and concerns Luma and Chroma components. Finally, \gls{lfnst} is enabled only when \gls{dct}-II is used as primary transform.       

\begin{figure}[t]
\includegraphics[width=0.48\textwidth]{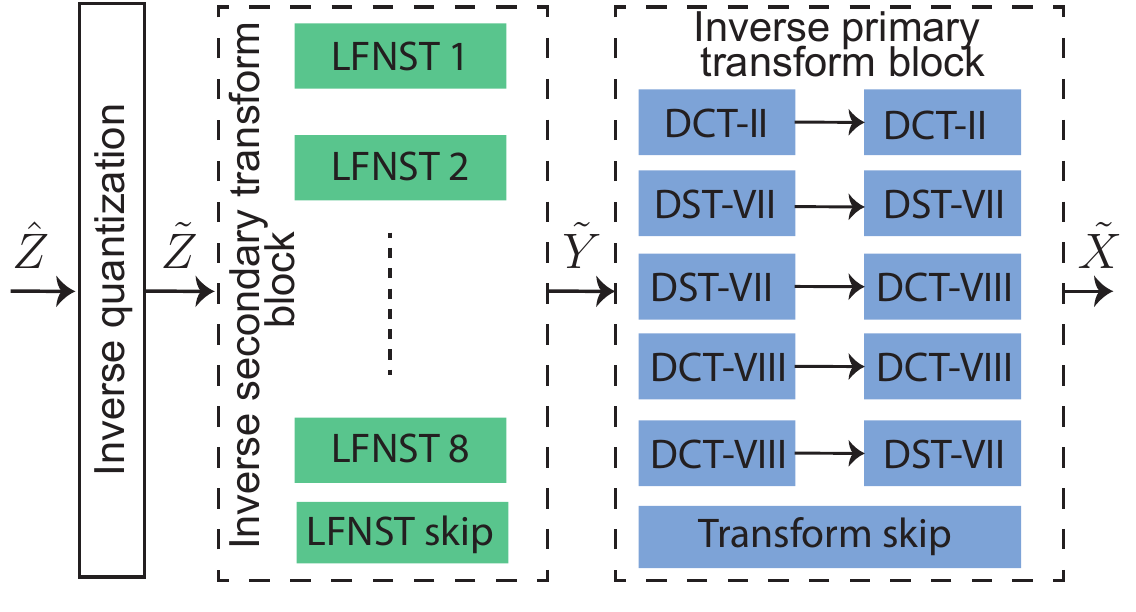}
\caption{Block diagram of the inverse VVC transform module}
\label{fig:vvc-trasform-module}
\end{figure}

To reduce the complexity in number of operations and memory required to store the transform coefficients, the $64\plh64$ inverse transform is reduced to $48 \plh 16$. Therefore, only 16 bases of the transform kernel are used and the number of input samples is reduced to 48 by excluding the bottom right $4\plh4$ block (ie. includes only samples of the top-left, top-right and bottom-left $4 \plh 4$ blocks). To further reduce the number of multiplications by sample, the \gls{lfnst} restricts the transform of $4\plh4$ and $8\plh8$ blocks to $8\plh16$ and $8\plh48$ transforms, respectively. In those cases, the \gls{lfnst} is applied only when the last significant coefficient is less than 8, and less than 16 for other block sizes. \Figure{\ref{fig:vvc-trasform-module}} illustrates the block diagram of the \gls{vvc} inverse transform module.   

\adcomment {The \gls{vvc} transform module raises many challenges to the hardware implementation of the \gls{vvc} decoder. The introduced \gls{dct}-II (p-64), \gls{dst}-VII, \gls{dct}-VIII kernels with the 8 \gls{lfnst} kernels will require high memory usage to store these coefficients. Moreover, these new transforms are more complex and require a higher number of multiplications as shown in Table~\ref{table::complex}. Finally, the \gls{lfnst} stage introduces an additional delay required to perform the secondary transform. Therefore, the hardware transform module should be carefully designed leveraging all optimisations to reach the target latency and throughput while minimising the hardware area.}     

\begin{table}[h]
\renewcommand{\arraystretch}{1.2}
\centering
\caption{Intra Prediction Mode (IPM) based  secondary transform signaling in VVC}
\label{table::LFNST_index}  
{ 
\begin{tabular}{c | c}
 \hline
\hline
Intra Prediction Mode & Transform set index \\
\hline
    $IPM<0$ & 1 \\ 
    $0 \leq IPM \leq 1$ & 0 \\ 
    $ 2 \leq IPM \leq 12$ & 1 \\ 
    $ 13 \leq IPM \leq 23$ & 2 \\ 
    $ 24 \leq IPM \leq 44$ & 3 \\
    $ 45 \leq IPM \leq 55$ & 2 \\
    $ 56 \leq IPM \leq 80$ & 1 \\
    $ 81 \leq IPM \leq 83$ & 0 \\
\hline
 \hline
\end{tabular}}
\end{table}

\section{Related Work}
In this section we give a brief description and analysis of existing hardware implementations of separable transforms proposed for \gls{hevc} and \gls{vvc}. 
\label{sec:related-works}
\subsection{Hardware Implementations of DCT-II}
\gls{dct}-II transform has been widely used by previous video coding standards such as \gls{avc}~\cite{1218189} and \gls{hevc}. Therefore, its hardware implementation has been well studied and optimized for different architectures. Shen { \it et al.}~\cite{6298499} proposed a unified \gls{vlsi} architecture for 4, 8, 16 and 32 point \gls{iict}. This latter relies on shared regular multipliers architecture that takes advantage of the recursion feature for large size blocks of 16- and 32-points \glspl{iict}. The proposed solution relies on a \gls{sram} to transpose the intermediate 1-D transform results which, in result, reduces logic resources. Zhu et {\it al.}~\cite{6571937} proposed a unified and fully pipelined 2D architecture for DCT-II, IDCT-II and Hadamard transforms. This architecture also benefits from hardware resources sharing and a \gls{sram} module to transpose the result of the intermediate 1D transform. Meher \textit{et al.}~\cite{6575105} proposed a reusable architecture for \gls{dct}-II supporting different transform sizes using constant multiplications instead of regular ones. This architecture has a fixed throughput of 32 coefficients per cycle regardless the transform size and it can be pruned to reduce the implementation complexity of both full-parallel and folded 2-D \gls{dct}-II. However, this approximation leads to a marginal effect on the coding performance varying from 0.8\% to 1\% of \gls{bdr} losses when both inverse \gls{dct}-II and \gls{dct}-II are pruned. 
Chen \textit{et al.}~\cite{Chen} proposed a 2D hardware design for the \gls{hevc} \gls{dct}-II transform. The presented reconfigurable architecture supports up to $32 \plh 32$ transform block sizes. To reduce logic utilization, this architecture benefits from several hardware resources, such as \gls{dsp} blocks, multipliers and memory blocks. The proposed architecture has been synthesized targeting different \gls{fpga} platforms showing that the design is able to encode 4Kp30 video at a reduced hardware cost.
Ahmed \textit{et al.}~\cite{Ahmed} proposed a dynamic N-point inverse \gls{dct}-II hardware implementation supporting all \gls{hevc} transform block sizes. The proposed architecture is partially folded in order to save area and speed up the design. This architecture reached an operating frequency of 150 MHz and supports real time processing of 1080p30 video. 

\subsection{Hardware Implementation of the MTS block}
Several works~\cite{Can-mert,Garrido, 8698857, Kammoun4x4, Kammoun, 9081944, 8794833} have recently investigated a hardware implementation of the earlier version of the \gls{mts} that includes several transform types. Mert \textit{et al.}~\cite{Can-mert} proposed a 2D hardware implementation including all five transform types for 4-point and 8-point sizes using additions and shifts instead of regular multiplications. This solution supports a 2D hardware implementation of the five transform types. However, transform sizes larger than 8 are not supported, which are more complex and would require more resources. A pipelined 1D hardware implementation for all block sizes from $4\plh4$ to $32\plh32$ was proposed by Garrido \textit{et al.}~\cite{Garrido}. However, this solution only considers 1D transform, while including the 2D transform  would normally be more complex.  Moreover, this design does not consider asymmetric block size combinations. This latter design has been extended by Garrido {\it et al.}~\cite{8698857} to support 2D transform using Dual port \gls{ram} as a transpose memory to store the 1D intermediate results. Authors proposed a pipelined 2D design placing two separate 1D processors in parallel for horizontal and vertical transforms. A multiplierless implementation of the \gls{mts} 4-point transform module was proposed by Kammoun \textit{et al.}~\cite{Kammoun4x4}. This solution has been extended to 2D hardware implementation of all block sizes (including asymmetric ones), by using the \gls{ip} Cores multipliers~\cite{LPM} to leverage the \glspl{dsp} blocks of the {\it Arria 10} platform~\cite{Kammoun}. This solution supports all transform types and enables a 2D transform process within an pipelined architecture. However, this design requires a high logic resources compared to solution proposed by Mert {\it et al.}~\cite{Can-mert} and Garrido {\it et al.} \cite{Garrido}. The approximation of the \gls{dst}-VII based on an inverse \gls{dct}-II transform and low complexity adjustment stage was investigated in~\cite{8907817}. This solution supports a unified architecture for both forward and inverse \gls{mts} with moderate hardware resources. However, the approximation introduces a slight bitrate loss while the architecture is not compliant with a \gls{vvc} decoder. Fan {\it et al.} ~\cite{8794833} proposed a pipelined 2D transform architecture for \gls{dct}-VIII and \gls{dst}-VII relying on the N-Dimensional Reduced Adder Graph (RAG-N) algorithm. The use of this algorithm enables an efficient use of adders and shifts to replace regular multipliers. However, this work does not include the support of \gls{dct}-II which requires high logic resources especially for large size block (ie. $64\plh64$). The only work found in the literature that supports all \gls{mts} types and sizes in 2-D was proposed in~\cite{9081944}. This work is an extension of the architectures proposed by Garrido \textit{et al.}~\cite{Garrido, 8698857}. This solution supports all the \gls{mts} block sizes up to $64 \plh 64$ including asymmetric blocks. However, the performance of the proposed design is considered to be low especially for a nominal scenario (worst case). The proposed architecture enables decoding UHD videos at 10 fps. Moreover, considering the worst case scenario, this solution is only able to decode HD videos at 32 fps. The main features and performance of the most aforementioned solutions are summarised in Tables~\ref{CompareMTS} and \ref{comparison2D} for \gls{fpga} and \gls{asic} platforms, respectively.

\section{Proposed lightweight hardware implementation of the VVC transform module}
\label{sec:proposed-solution}

\begin{figure}[hbt!] 
\centering
\includegraphics[width=1\linewidth]{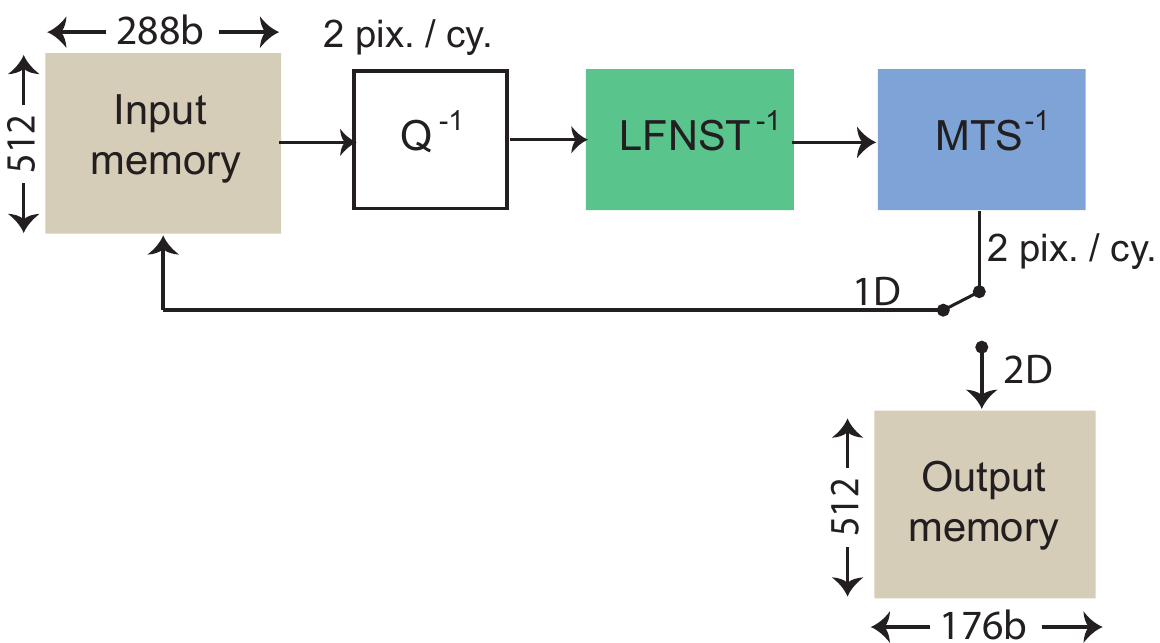}
\centering
\caption{Inverse quantization and transform block diagram}
\label{fig:top_levl}
\end{figure}

In this section, our proposed hardware designs for the \gls{vvc} transform process: inverse \gls{mts} and inverse \gls{lfnst} are described. Both modules were designed so that they can use the same input and output memories. \Figure{\ref{fig:top_levl}} depicts the top level hardware architecture that regroups the separable and non-separable transforms. The module targets a throughput of 1 sample/cycle for the entire \gls{vvc} transform. For that, the 1-D \gls{mts} was designed on the measure of 2 samples/cycle. As a result, we get a mean throughput of 1 sample/cycle for the 2-D \gls{mts}. Since the Inverse quantisation and the Inverse \gls{lfnst} share with the 1-D Inverse \gls{mts} the same input/output memories, they follow the same throughput of 2 samples/cycle as shown in \Figure{\ref{fig:top_levl}}. To further enhance chaining between transform blocks, the modules sustain a fixed system latency ($L1$ and $L2$ for inverse \gls{lfnst} and inverse \gls{mts} respectively). This enables an exact prediction of the \gls{vvc} transform performance. The transform modules follow the latency of the $64 \plh 64$ transform block, regardless of the input block size. In fact, 1 sample/cycle throughput at a target operating frequency of 600 MHz enables the processing of more than 35 frames per second for $3840 \plh 2160$ (4K) resolution videos in 4:2:2 Chroma sub-sampling.

When the first direction is selected and the \gls{dct} of type II is chosen for the \gls{mts}, the \gls{lfnst} is enabled and data will be modified by this module. Otherwise, when the second direction or the \gls{dct} type VIII or \gls{dst} type VII is selected for the \gls{mts}, the \gls{lfnst} is disabled and data will go through the module unchanged but by maintaining its latency with the same throughput. 
The input and output memories are designed to hold a \gls{ctu} of size $64 \plh 64$ for a 4:2:2 Chroma sub-sampling. As a result, the memory has been designed to hold one Luma \gls{ctb} of size $64 \plh 64$ and two Chroma \gls{ctb}s (Cr and Cb) of sizes $64 \plh 32$. The input memory has two main roles: it stores the transform samples and it is used as a transpose memory, therefore, the sample size is 18-bits (bounded by the maximum length of \gls{hevc} transform coefficients). The output memory stores only residuals which makes its sample size of 11-bits. Both memories have 512 lines, with 288-bits as depth for the input memory which results in a total size of 18.432 Kbytes, and 176-bits depth for the output which makes the total output memory size 11.264 Kbytes.

\begin{figure*}[t]
\includegraphics[width=1 \textwidth]{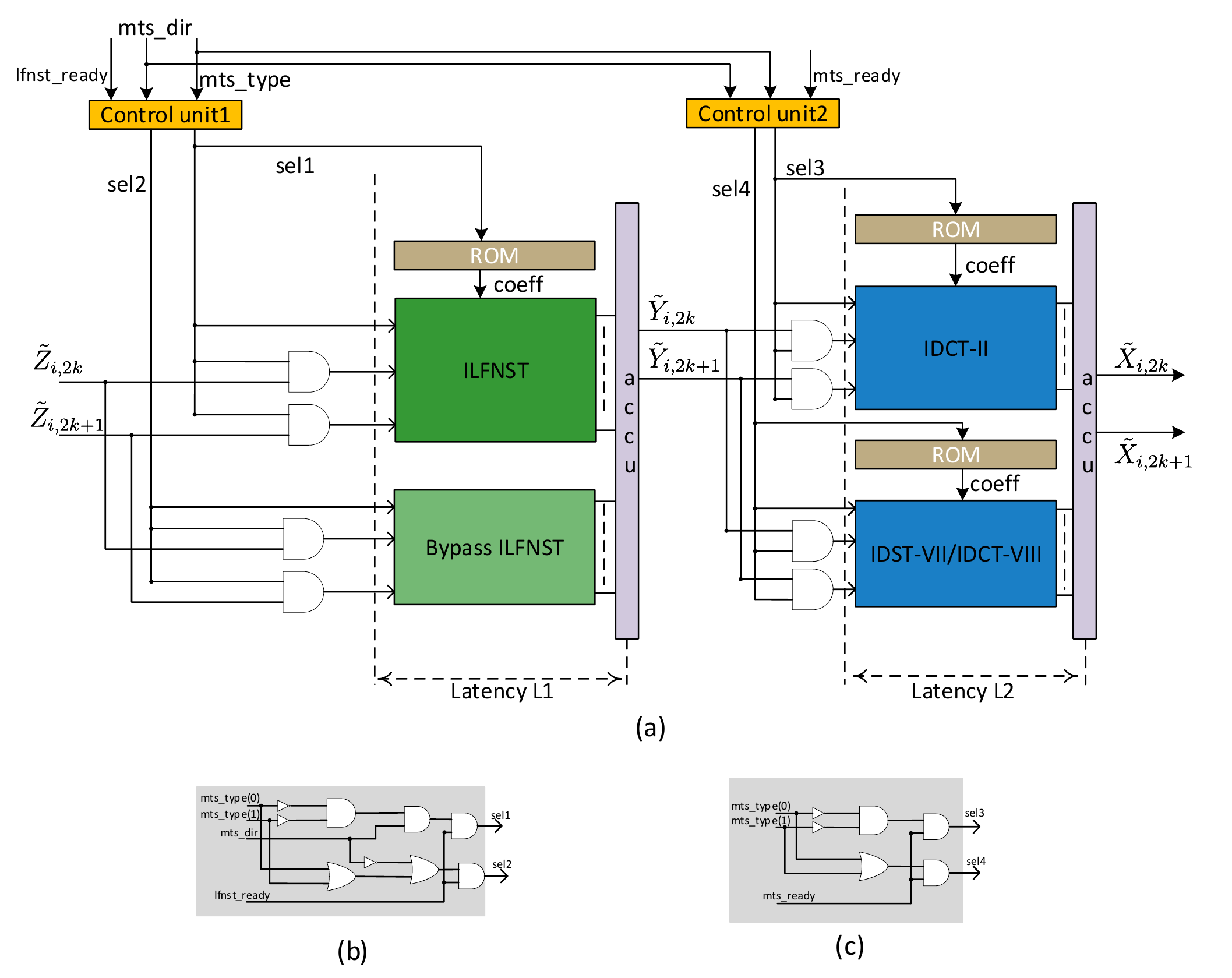}
\caption{Proposed hardware \gls{mts} and \gls{lfnst} architecture, (a) Unified design for all sizes and all transform types, (b) Control unit 1 for \gls{mts}, (c) Control unit 2 for \gls{lfnst}.}
\label{fig:mts_lfnsttplevl}
\end{figure*}

\adcomment{\Figure{\ref{fig:mts_lfnsttplevl}} illustrates the proposed hardware architecture for inverse \gls{mts} and \gls{lfnst} modules. The first control unit,  shown in \Figure{\ref{fig:mts_lfnsttplevl}}:(b), outputs two selection signals {\it sel1} and {\it sel2} which are derived based on the transform type {\it mts\_type} (ie. DCT-VII or DCT-VIII/DST-VII) and the direction {\it mts\_dir} (ie. vertical or horizontal). If the transform type ({\it mts\_type}) refers to \gls{dct}-II, the input samples are processed by the inverse \gls{lfnst} module, otherwise they go through the { \it Bypass} module. The inverse \gls{lfnst} module uses 32 multipliers in a shared architecture for all block sizes and kernels. Both inverse \gls{lfnst} and bypass modules have the same latency {\it L1}, for that, the {\it Bypass} module uses registers to create a delay line with latency {\it L1}. At the end of the inverse \gls{lfnst}, the output samples are accumulated and delivered to the \gls{mts} modules in a 2 samples/cycle rate. The second control unit, shown in \Figure{\ref{fig:mts_lfnsttplevl}}:(c), delivers two signals {\it sel3} and {\it sel4}. These signals are derived from the transform type {\it mts\_type}. {\it sel3} enables the \gls{dct}-II module, while {\it sel4} enables \gls{dct}-VIII/\gls{dst}-VII module. These two modules share 32 multipliers and have the same latency {\it L2}. At the end of the transformation, a couple of output samples are delivered through $\widetilde X_{i,2k}$ and  $\widetilde X_{i,2k+1}$ signals every clock cycle $k$.}

\subsection{Hardware Separable transform module} 

The \gls{mts} hardware architecture was built based on a constraint of 2 samples/cycle throughput and a fixed latency. The architecture of 4/8/16/32-point \gls{dct}-II/VIII, \gls{dst}-VII and 64-point \gls{dct}-II uses 32 \gls{rm}. Thirty-two is the minimum number of multipliers needed to get a throughput of 2 samples/cycle. This number is bounded by the odd butterfly decomposition of the $64\plh64$ \gls{dct}-II transform matrix and the $32\plh32$ \gls{dct}-VIII/\gls{dst}-VII transform matrices.     

The zeroing concept in \gls{vvc} applied on large block sizes enable processing one sample/cycle at the input to get a 2 sample/cycle at the output. In fact, for 64-point I\gls{dct}, the size of the input vector is 32 and the output vector size is 64. Concerning the 32-point I\gls{dct}-VIII/I\gls{dst}-VII, the size of the input and output vectors are 16 and 32, respectively. In both cases, the output vector is twice the size of the input vector enabling to lower the input rate to one sample per cycle. 

\begin{figure}[hbt!] 
\centering
\includegraphics[width=1\linewidth]{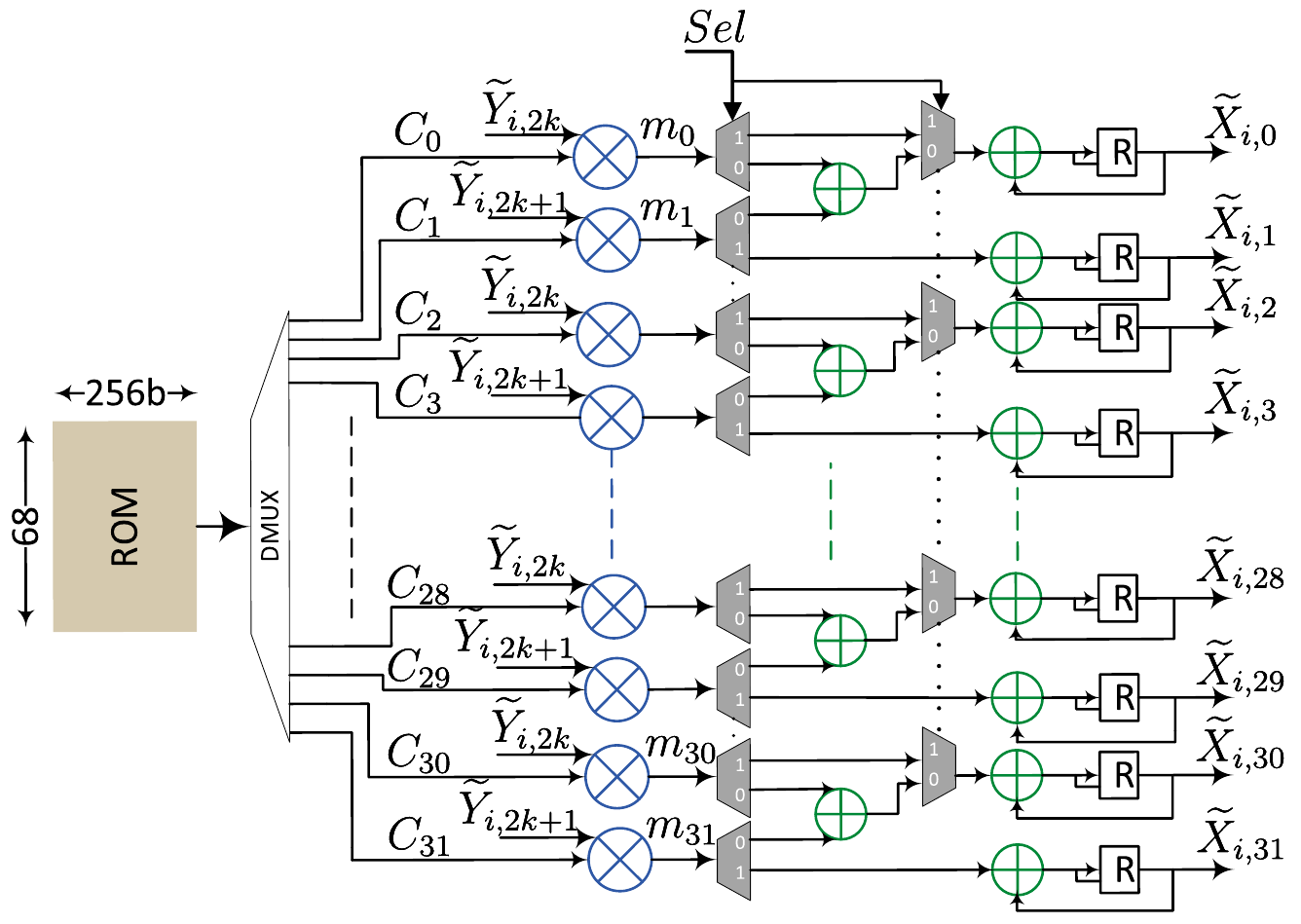}
\centering
\caption{\gls{mts} \gls{rm} architecture, $\tilde{Y}_{i, 2k}$ and $\tilde{Y}_{i, 2k+1}$ are the input samples loaded from the $input$ bus at clock cycle $k$. $C$ is the coefficient vector, $m_i$ are the \glspl{rm} and $\tilde{X}_{i, j}$ are samples of the output vector}
\label{fig:shared_mult}
\end{figure}

\Figure{\ref{fig:shared_mult}} shows the architecture of the hardware module using 32 \glspl{rm} referred to $m_i$ $i \in \{0, \dots, 31 \}$. $\tilde{Y}_{i, 2k}$ and $\tilde{Y}_{i, 2k+1}$ are the input samples at clock cycle $k$. Using the zeroing for large block sizes, $\tilde{Y}_{i, 2k}$ and $\tilde{Y}_{i, 2k+1}$ carry the same input sample. Otherwise, they carry two different samples. The input samples are then multiplied by the corresponding transform coefficients vector $C_i$  $(i \in {0, \dots, N-1})$, where $C$ holds a line of the transform matrix in case of zeroing and two lines otherwise. The output results are then accumulated in the output vector $\tilde{X}_i$. This accumulation is depicted in the figure through the adders and the feedback lines. 

The transform coefficients are stored in a \gls{rom}. The total memory size is 17408-bits which corresponds to 68 columns of 256 bit-depth (256$\plh$68). The \gls{rom} stores the coefficients of the 64-point \gls{dct}-II, 32-point, 16-point, 8-point and 4-point \gls{dst}-VII matrix coefficients. The 64-point \gls{dct}-II is decomposed using its butterfly structure, and the resultant sub-matrices are stored. Moreover, one sub-matrix (16-point) is replicated in order to respect the output rate. It is well known that IDCT-VIII $C_8^T$ can be computed from the IDST-VII $S_7^T$ using pre-processing $\Lambda$ and post-processing $\Gamma$ matrices as expressed in
~(\ref{Eq9})~\cite{6638744}. 
\begin{equation}
\left[ C^N_{8} \right]^T  = \Lambda^N \cdot \left[ S^N_{7}\right]^T \cdot \Gamma^N,   
\label{Eq9}
\end{equation}
where $\Lambda$ and $\Gamma$ are permutation and sign changes matrices computed in~(\ref{Eq5}) and ~(\ref{Eq6}), respectively.
\begin{equation}
\Lambda_{i,j}^N= \left\{ \begin{array}{cc}1, & \textit{ if } j=N-1-i, \\ 0, & \text{ otherwise }, \end{array}\right. \\
\label{Eq5}
\end{equation}
\begin{equation}
\Gamma_{i,j}^N=\left\{ \begin{array}{cc} (-1)^{i}, & \textit{ if } j=i,  \\ 0, & \text{ otherwise }, \end{array}\right. \\ 
\label{Eq6}
\end{equation}
with $i, j \in \{0, 1, \dots,  N-1\}$ and $N \in \{4, 8, 16, 32 \}$.

\begin{figure}[hbt!] 
\centering
\includegraphics[width=1\linewidth]{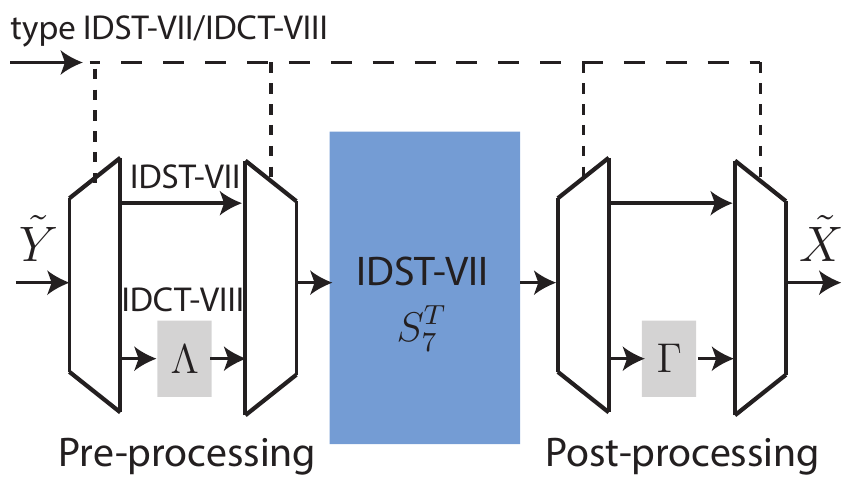}
\centering
\caption{Unified computing of inverse \gls{dst}-VII and inverse \gls{dct}-VIII}
\label{fig:dst7_dct8}
\end{figure}

This relationship between \gls{dct}-VIII and \gls{dst}-VII enables to compute both transforms using one kernel. Therefore, we store only the \gls{dst}-VII transform matrices. \Figure{\ref{fig:dst7_dct8}} illustrates the unified design to process both \gls{dst}-VII and \gls{dst}-VIII. The post and pre-processing matrices do not require any additional multiplication since they are composed of only $1$ and $-1$ involving sign changes and permutation operations. 

\subsection{Hardware Non-Separable transform module}

The non-separable transform or the \gls{lfnst} was designed on the measure of the 2 samples/cycle throughput. The \gls{lfnst} architecture supports all \gls{vvc} transform block sizes including asymmetrical blocks. As discussed in Section~\ref{sec:lfnst}, the \gls{lfnst} operates on the $4 \plh 4$ top-left corner of the input block. However, depending on the input block width and height, the 16 residuals of the $4 \plh 4$ corner may not all be used. For example, for the 4$\plh$4 block only 8 residuals of the 16 are transformed. The output of the \gls{lfnst} depends on the input block width and height, which can be either 16 or 48. The 4$\plh$4 top-left corner is zigzag scanned and transformed into a line vector. This latter is then fed to the \gls{lfnst} module. Table~\ref{tab:lfnst_in_out} presents the input/output sizes of the \gls{lfnst} core according to the transform block size. 
\begin{table}[hbt!]
\renewcommand{\arraystretch}{1.2}
\caption{Core \gls{lfnst} input and output sizes}
\centering
\begin{tabular}{l|c|c}
	\hline
		\hline
	Block size & Input size  & Output size  \\
	\hline
	$4 \plh 4$ & 8& 16\\	
	
	$8\plh 8$ & 8 & 48\\
	
	$4 \plh N$ and $M \plh 4 $ ($N, M>4$) & 16 & 16\\

	$8 \plh N$ and $M \plh 8$ ($N, M>8$) & 16 & 48\\
	\hline
	 \hline
\end{tabular}
\label{tab:lfnst_in_out}
\end{table}


We can notice from Table~\ref{tab:lfnst_in_out} that due to the difference between the input and output sizes, their corresponding throughput can also be different. To perform constant throughout for the 4$\plh$4 block, we can lower the input rate to 1 sample per 2 clock cycles. For 8$\plh$8 block sizes, we can lower it to 1 sample per 3 clock cycles. For $8 \plh N$ and $M \plh 8$ blocks, the input rate is set to 2 samples per 3 cycles. Concerning the $4 \plh N$ and $M \plh 4$ ($N, M > 4$) blocks, the input/output rates remain the same, 2 samples/cycle. In all cases, we get the desired output rate of 2 samples/cycle. More details in input/output rates are given in Table~\ref{tab:io-rate} for different block sizes. 
\begin{table}[hbt!]
\renewcommand{\arraystretch}{1.2}
\caption{\gls{lfnst} input/output throughput depending on block size.}
\centering
\begin{tabular}{l|c|c}
	\hline
		\hline
\multirow{2}{*}{Block size}	 & Input rate  & Output rate  \\
	 &  (sample/cycle) &  (sample/cycle) \\
	\hline
	$4 \plh 4$ & 1& 2\\	
	$8\plh 8$ & 1/3 & 2\\
	$4 \plh N$ and $M \plh 4 $ ($N, M>4$) & 2 & 2\\
	$8 \plh N$ and $M \plh 8$ ($N, M>8$) & 2/3 & 2\\
	\hline
		\hline
\end{tabular}
\label{tab:io-rate}
\end{table}

To perform the secondary transform, \gls{lfnst} uses a total of 16 kernels, 8 kernels of size $16 \plh 16$ and 8 others of size $16 \plh 48$. These kernels are stored in a \gls{rom}. The total \gls{rom} size is 65536-bits, which corresponds to 256 columns of 256-bits in depth ($256 \plh 256$). Like the \gls{mts}, the proposed \gls{lfnst} design uses 32 \gls{rm}, which is the minimum number of multipliers needed to satisfy the target 2 samples/cycle rate. As a result, the \gls{lfnst} core design follows the same design as the \gls{mts} depicted in \Figure{\ref{fig:shared_mult}}, but unlike the \gls{mts}, \gls{lfnst} adds a delay line at the output.

\begin{figure}[hbt!] 
\centering
\includegraphics[width=1\linewidth]{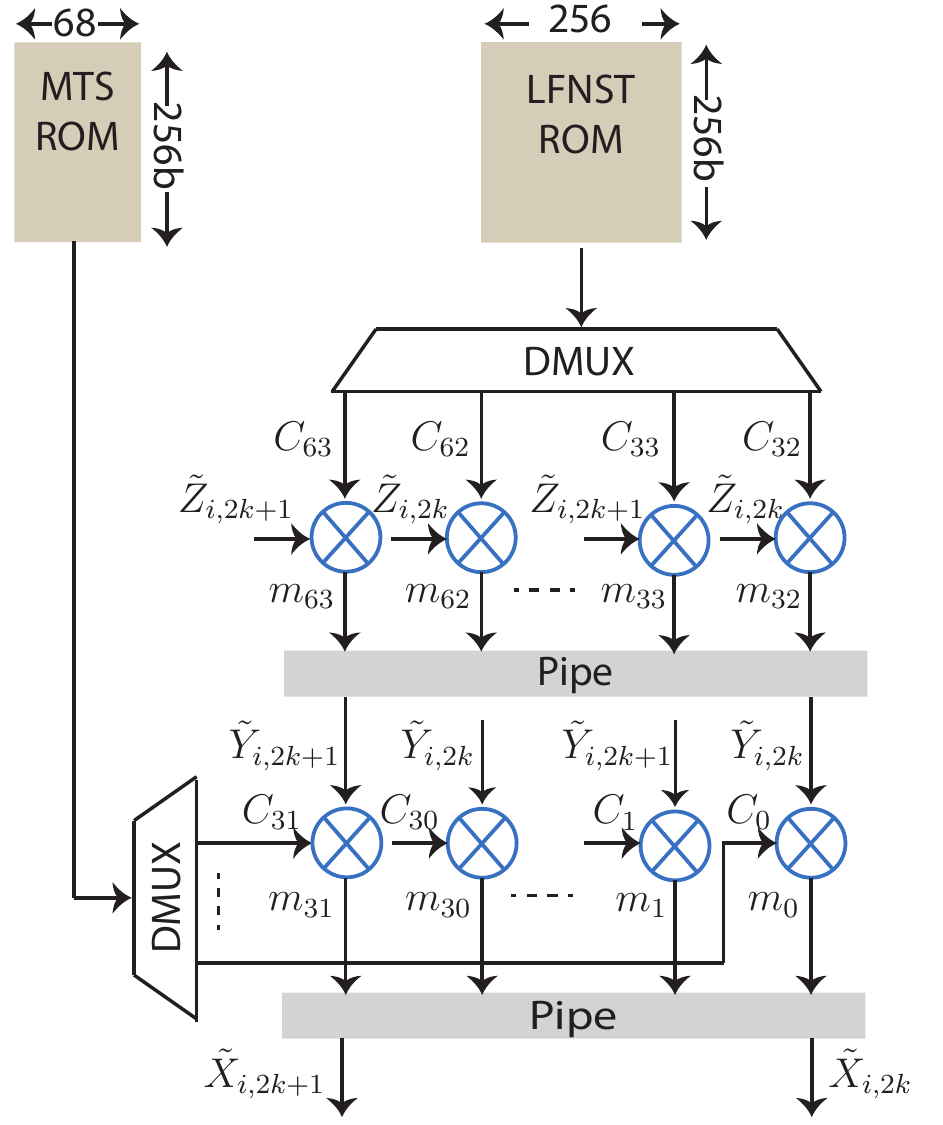}
\centering
\caption{Hardware architecture of the \gls{vvc} transform module including \gls{mts} and \gls{lfnst}. Two samples $\tilde{X}_{i, 2k}$ and $\tilde{X}_{i, 2k+1}$ are processed each clock cycle $k$}
\label{fig:shared_mult2}
\end{figure}

\Figure{\ref{fig:shared_mult2}} depicts the VVC transform (\gls{mts} + \gls{lfnst}) unified design. The proposed designs use in total 64 \gls{rm} $m_j$, 32 for the \gls{mts} $j \in \{0, \dots, 31 \}$ and 32 for the \gls{lfnst} $j \in \{32, \dots, 63 \}$. As we can see in the \Figure{\ref{fig:shared_mult2}}, the \gls{lfnst} is performed before the \gls{mts}, from its \gls{rom} it retrieves, depending on the \gls{lfnst} index and set index, the 32 coefficients that correspond to the input residuals shown as $C_j$ with $j \in \{32, \dots, 63 \}$. These coefficients are then fed to the core multipliers along with the input samples $\tilde{Z}_{i, 2k}$ and $\tilde{Z}_{i, 2k+1}$ where $k$ is the clock cycle. The result of the multiplication is forwarded to the delay line which is depicted by the pipe in the figure. This latter ensures a constant system latency for the \gls{lfnst}. At the end of the \gls{lfnst} processing, the output samples feed the \gls{mts} module with a rate of 2 samples/cycle through $\tilde{Y}_{i, 2k}$ and $\tilde{Y}_{i, 2k+1}$ signals. These samples are then fed to the core multipliers along with the corresponding coefficients retrieved from the \gls{mts} \gls{rom} $C_j$ with $j \in \{0, \dots, 31 \}$. The final result goes to the second delay line which unifies the latency for the \gls{mts}. Finally, the transformed samples are redirected to either the transpose or output memory via $\tilde{X}_{i, 2k}$ and $\tilde{X}_{i, 2k+1}$ signals. They are directed to the transpose memory if the first direction is selected for the \gls{mts} and to the output memory otherwise.

The \gls{lfnst} + \gls{mts} processor control interface is summarized in Table \ref{tab:Tr-interface}. A positive pulse in $input\_enable$ launches the transform process, with the block size defined by {\it tr$\_$width} and {\it tr$\_$height}. The \gls{mts} and \gls{lfnst} control signals are defined in the interface with prefix $MTS\_..$ and $LFNST\_..$, respectively. For example, the transform type and direction for the \gls{mts} are defined in the {\it MTS$\_$type} and {\it MTS$\_$dir} input signals, respectively. Following the {\it input$\_$enable} the $tr\_in$ signal will carry two $N_{bi}$ samples at the next clock cycle. These samples are then fed to the \gls{lfnst} module. The result, carried by {\it LFNST$\_$out}, is then sent to the \gls{mts} module, which will perform a 2 dimensional transform. After finishing the 2-D transform, the \gls{mts} communicates the transformed samples to the output memory via {\it MTS$\_$out$\_$fin} signal. However, if the first direction is selected for the \gls{mts}, the results will be directed to the input memory via {\it MTS$\_$out$\_$inter}. For both modules that integrate data loading and pipeline stages, the design starts generating the results of the \gls{lfnst}/1-D \gls{mts} after a fixed system latency. It then generates thanks to the piplined architecture the outputs every clock cycle without any stall.

\begin{table}[hbt!]
\renewcommand{\arraystretch}{1.2}
\caption{\gls{lfnst} design interface.}
\centering
\tabcolsep=0.10cm
\begin{tabular}{l|c|c|l}
	\hline
		\hline
	Signals & I/O & \#BITS & Description \\
	\hline
	$clk$ & I& 1& System Clock\\	
	{\it rst$\_$n} & I & 1 & System reset, active low\\
	{\it  input$\_$enable } & I& 1& { Activation pulse to start}\\
	{\it  AVC$\_$VVC} & I& 1& { Video standard: 0, AVC; 1, HEVC or VVC} \\
	{\it  tr$\_$width} & I& 3& {Transform Block width: 000:4 - 100:64} \\
	{\it  tr$\_$height} & I& 3& {Transform Block height: 000:4 - 100:64}\\
	{\it MTS$\_$type} & I& 2& {Tran. type: DCT-II, DCT-VIII and DST-VII} \\
	{\it  MTS$\_$dir}	& I & 1 & {Transform direction : 0: Hori.; 1: Vert.}\\
	{\it  LFNST$\_$pos$\_$x} & I & 5 & {Position x of vector input}\\
{\it  	LFNST$\_$pos$\_$y}	& I & 3 & {Position y of vector input}\\
	{\it  LFNST$\_$set$\_$idx} & I & 2 & LFNST kernel set index \\
{\it 	LFNST$\_$idx} & I & 2 & LFNST kernel index \\
	{\it  tr$\_$src$\_$in} & I & 2$\, N_{bi}$ & {\gls{lfnst} input data}\\
{\it 	MTS$\_$ready} & O & 1 & {Ready pulse, end of N-point \gls{mts}}\\
	{\it  LFNST$\_$ready} & O & 1 & {Ready pulse, end of LFNST}\\
{\it 	MTS$\_$out$\_$inter} & O & 2$\, N_{bi}$ & Intermediate output, 1-D \gls{mts} \\
	{\it  MTS$\_$out$\_$fin} & O & 2$\, N_{bo}$ & Final output, 2-D \gls{mts} \\
{\it 	LFNST$\_$out} & O & 2$\, N_{bi}$ & LFNST output data\\
	\hline
    \hline
\end{tabular}
\label{tab:Tr-interface}
\end{table}

\begin{table*}
\centering
\renewcommand{\arraystretch}{1.2}
\caption{Performance ($\%$) in terms of \gls{bdr} and run time complexity when \gls{mts} and \gls{lfnst} tools turned off in VTM8.0~\cite{JVET-R0013-v3}. \adcomment{Evaluations performed under the VVC common test conditions~\cite{JVET-H1010}}.}
\label{table::bd-rate}  
\begin{tabular}{l | c c c c c | c c c c c | c c c c c }
\hline
\hline
Disabled tool & \multicolumn{5}{c|}{All Intra Main 10} & \multicolumn{5}{c|}{Random Access Main 10} & \multicolumn{5}{c}{Low Delay B Main 10}  \\ 
 \hline
 & Y & U & V  & \textit{EncT} & \textit{DecT} & Y & U & V & \textit{EncT} & \textit{DecT} & Y  & U & V & \textit{EncT} & \textit{DecT}  \\ 
\hline
 \gls{mts} &  $1.32$  &  $0.96$& $1.02$& $85$ & $99$ & $0.75$ & $0.60$ & $0.53$ & $89$ & $95$ & $0.53$ & $0.26$ & $0.04$ & $93$ & $96$   \\
\gls{lfnst} &  $0.99$  &$1.98$ &$2.21$ & $110$ & $100$ & $0.70$ &$0.78$ & $1.08$& $96$ & $100$ & $0.33$ &$0.88$& $0.96$& $107$ & $98$   \\
\hline
\hline
\end{tabular}
\label{tab:coding-efficiency}
\end{table*}

\section{Experimental Results}
\label{sec:results}    

\subsection{Experimental Setup}

VHDL hardware description language is used to describe the proposed transform block. A state-of-the-art logic simulator \cite{RiviraPro} is used to test the functionality of the \gls{vvc} transform module (\gls{mts} + \gls{lfnst}). The design was synthesised by a commercial Design Compiler using 28-nm technology. The test strategy is performed as follows. First a set of $10^5$ pseudo-random input vectors have been generated and used as test patterns to test the 1-D \gls{mts} in unit tests. The same number of tests was used to verify the \gls{lfnst} module. Second, a software implementation of the inverse \gls{mts} and \gls{lfnst} has been developed, based on the transform procedures defined in the Versatile Video Coding Test Model (VTM) version 8.0
~\cite{JVET-Q2002-v2}. Using self-check technique, the bit accurate test-bench compares the simulation results with those obtained using the reference software implementation. The tests cover all transform block sizes from $4 \plh 4$ to $64 \plh 64$ including asymmetric blocks, for both \gls{mts} and \gls{lfnst} modules.  

\subsection{Results and Analysis}

\subsubsection{Coding performance}
First the coding gain and  software complexity of the \gls{mts} and \gls{lfnst} modules under the VVC common Test Conditions in three main coding configurations including All Intra, Random Access and Low Delay B are analysed~\cite{CTCVVC2018}. \adcomment{Table~\ref{tab:coding-efficiency}~\cite{JVET-R0013-v3} gives the coding losses and the complexity reductions at both encoder and decoder when the \gls{mts} and \gls{lfnst} tools are turned off. We can notice that disabling \gls{mts} and \gls{lfnst} tools introduces the highest coding losses in All Intra coding configuration with 1.32\% and 0.99\% \gls{bdr}
~\cite{bjontegaard2001calculation, bjontegaard2008improvements} losses when these two tools are disabled, respectively.} The \gls{mts} and \gls{lfnst} tools are active only for Intra blocks to limit the complexity overhead at the encoder side. This may explain the higher coding gain brought by the \gls{mts} and \gls{lfnst} in All Intra configuration compared to the two other Inter configurations. Moreover, the \gls{mts} tool significantly increases the encoder complexity since five sets of horizontal and vertical transforms are tested to select the best performing set of transforms in terms of rate distortion cost. At the decoder side, the \gls{mts} slightly increases the decoder complexity since only one transform set is processed by block. This slight complexity overhead is mainly caused by introducing more complex transforms including \gls{dst}-VII, \gls{dct}-VIII and \gls{dct}-II of size 64. Disabling the \gls{lfnst} tool changes the encoder behaviour resulting in complexity increase at the encoder side caused by complexity reduction tools included in the VTM to speedup the encoder such as early termination~\cite{8803533, 8826595}. \adcomment{In fact, the \gls{lfnst} tool enables more efficient coding of residuals resulting in better energy packing and more coefficients are set to zeros after the quantisation. Therefore, the early termination techniques integrated in the VTM~\cite{8803533} terminate the RDO process earlier when the \gls{lfnst} is enabled resulting in lower encoding time.}  At the decoder side, disabling the \gls{lfnst} slightly decreases the decoder complexity since \gls{lfnst} is applied only on blocks processed by the \gls{dct}-II primary transform. We can conclude from this study that both \gls{mts} and \gls{lfnst} have a slight impact on the software decoder complexity. In the next section, the impact of both \gls{mts} and \gls{lfnst} tools will be investigated on a hardware decoder targeting an \gls{asic} platform.             

\subsubsection{MTS performance}

The proposed 1-d \gls{mts} design was synthesized using Design Compiler (DC) targeting an 28-nm  \gls{asic} at 600 MHz. The design sustain 2 samples/cycle with a fixed system latency. The total area consumed by the 1-D \gls{mts} is 87.7 Kgates. The 2-D \gls{mts} is applied using the 1-D core in a folded architecture, where the output of the first direction is redirected to a transpose memory which re-sends the transposed data to the same 1-D core. As a result the 2-D \gls{mts} includes both 1-D \gls{mts} and the transpose memory. The same architecture can be found in many state of the art works, however the proposed \gls{mts} design supports all sizes and types of \gls{vvc} transform including the block of order 64 for the \gls{dct}-II. This design proved to be scalable enabling to doubled the performance of the design from a throughput of 2 samples/cycle to a throughput of 4 samples/cycle. The flexibility of the design allowed this transition to be done only by doubling the total number of multipliers from 32 to 64 multipliers while conserving the same architecture. Table \ref{tab:synASIC1} shows the synthesis results for both design, the 4 samples/cycle design increases the area cost by 55\% instead of 100\% and this is because doubling the number of multipliers reduces the size of the delay line which is not negligible compared to the total area size.
\begin{table}[!h]
\renewcommand{\arraystretch}{1.2}
\caption{Synthesis results for \gls{mts}-2 p/c and 4 p/c at 600 MHz.}
\centering
\begin{tabular}{l|l|c|c}
	\hline
	\hline
 	& & MTS-2p/c  & MTS-4p/c \\
	\hline
\multirow{4}{*}{\gls{asic} 28-nm}	& Num. of mult. & 32 & 64 \\
	& Combinational area & 59135 & 97210 \\
	& Non-combinational area & 29946 & 41705 \\
	& Total area (gate count) & 89082 & 138916 \\
	\hline
	\hline
\end{tabular}
\label{tab:synASIC1}
\end{table}

\begin{table}[!h]
\renewcommand{\arraystretch}{1.2}
\caption{Synthesis results for \gls{lfnst}-2 p/c and 4 p/c at 600 MHz.}
\centering
\begin{tabular}{l|l|c|c}
	\hline
		\hline
 	& & LFNST-2p/c  & LFNST-4p/c \\
	\hline
\multirow{4}{*}{\gls{asic} 28-nm}	& Num. of mult. & 32 & 64 \\
	& Combinational area &  49911 & 79409 \\
	& Noncombinational area & 24680 & 29806 \\
	& Total area (gate count)& 74592 & 109215 \\
	\hline
		\hline
\end{tabular}
\label{tab:synASIC2}
\end{table}

\begin{table}[!h]
\renewcommand{\arraystretch}{1.2}
\caption{VVC vs HEVC Transform synthesis results at 600 MHz.}
\centering
\begin{tabular}{l|l|c|l}
	\hline
		\hline
 	& & VVC  & HEVC  \\
	\hline  
   \multirow{2}{*}{\gls{asic} 28-nm}	
	& Architecture & RM & MCM \\
	& Total area & 163672 & 41500 \\
	\hline
		\hline
\end{tabular}
\label{tab:synVVC-HEVC}
\end{table}

\begin{table}[!h]
\centering
\renewcommand{\arraystretch}{1.2}
\caption{\adcomment{Comparison of \gls{mts} hardware designs on \gls{fpga} platform}.}
\vspace{1mm}
\begin{tabular}{p{0.9in}|p{1in}|p{1in}} \hline 	\hline
Solutions&   \centering{Garrido {\it et al.}~\cite{9081944}} & \centering{Proposed architecture}\tabularnewline \hline 
 Technology  &  \centering{ME 20 nm \gls{fpga}} & \centering{Arria 10 \gls{fpga}}  \tabularnewline  
 ALMs   &  \centering{5179}&  \centering{9723} \tabularnewline 
 Registers   &  \centering{9104}&  \centering{14368}  \tabularnewline
 DSPs   & \centering{32} &  \centering{32} \tabularnewline 
Frequency (MHz)   &   \centering{{204}}&  \centering{165}\tabularnewline 
Throughput (fps) &  \centering{1920 $\plh$ 1080p40}&  \centering{1920$\times$1080p50} \tabularnewline 
Memory  &  \centering{3910Kb}&  \centering{$-$}\tabularnewline 
\gls{mts} size  &     \centering{4, 8, 16, 32, 64}  &  \centering{4, 8, 16, 32, 64}\tabularnewline  
 \multirow{2}{*}{\gls{mts} type}	  & \centering{ \gls{dct}-II/VIII, \gls{dst}-VII}  & \centering{\gls{dct}-II/VIII, \gls{dst}-VII} \tabularnewline 
\hline
\hline 
\end{tabular}
\label{CompareMTS}
\end{table}

\begin{table*}[!h]
\centering
\renewcommand{\arraystretch}{1.2}
\caption{Comparison of different hardware transform designs on \gls{asic} platform.}
\vspace{1mm}
\begin{tabular}{p{1.0in}|p{1.2in}|p{1.2in}|p{1.2in}|p{1.2in}} \hline 	\hline
Solutions&  \centering{Jia {\it et al.}~\cite{6571937}}  & Mert {\it et al.} \centering{~\cite{Can-mert}} &  Fan {\it et al.} \centering{~\cite{8794833}}& \centering{Proposed architecture}\tabularnewline \hline 
 Technology  &  \centering{\gls{asic} 90 nm} &  \centering{\gls{asic} 90 nm} &  \centering{\gls{asic} 65 nm}& \centering{\gls{asic} 28 nm}  \tabularnewline  
 Gates   &  \centering{235400} &  \centering{ 417000} &  \centering{496400}&\centering{163674} \tabularnewline 
Frequency (MHz)   &  \centering{311} &   \centering{160} &  \centering{250} & \centering{600}\tabularnewline 
Throughput (fps) &  \centering{1920 $\plh$ 1080p20} &   \centering{7680 $\plh$ 4320p39} &  \centering{$-$} & \centering{3840$\plh$2160p30} \tabularnewline 
Memory  &  \centering{$108,1$ Kb} &  \centering{$-$} &  \centering{$-$} & \centering{$147,456$ Kb}\tabularnewline 
\gls{mts} size  &  \centering{ 4, 8, 16, 32}  &     \centering{ 4, 8} &   \centering{4, 8, 16, 32}  &  \centering{4, 8, 16, 32, 64}\tabularnewline  
\gls{mts} type & \centering{ {\gls{dct}-II, Hadamard} } &   \centering{ \gls{dct}-II/VIII, \gls{dst}-VII} &  \centering{\gls{dct}-VIII, \gls{dst}-VII} & \centering{\gls{dct}-II/VIII, \gls{dst}-VII} \tabularnewline 
\gls{lfnst}   &  \centering{\XSolidBrush} &   \centering{\XSolidBrush} & \centering{\XSolidBrush} & \centering{\Checkmark}\tabularnewline
\gls{lfnst} size  &  \centering{$-$} &  \centering{$-$} & \centering{$-$} & \centering{$4 \plh 4$ to $64 \plh 64$}\tabularnewline 
	\hline
\hline 
\end{tabular}
\label{comparison2D}
\end{table*}
\subsubsection{LFNST performance}

Like the \gls{mts} the proposed \gls{lfnst} design was synthesized using Design Compiler (DC) targeting an 28-nm \gls{asic} at 600 MHz. Because it is chained to the 1-D \gls{mts} it sustains the same throughput of 2 samples/cycle with a fixed system latency. The total area consumed by the \gls{lfnst} design is 71.6 Kgates. Unlike the \gls{mts}, the \gls{lfnst} is a new tool introduced in the \gls{vvc} standard. That is why no study was found in literature that investigates a hardware implementation of \gls{lfnst}. As a result, this work is the first study for a hardware architecture for the \gls{lfnst}. Similar to \gls{mts}, the scalability of \gls{lfnst} design enabled a very flexible transition going from a throughput of 2 samples/cycle to a throughput of 4 samples/cycle. The design proved to be scalable, doubling the throughput comes only at the expense of doubling the total number of multipliers while the same architecture is preserved. Table \ref{tab:synASIC2} shows the synthesis results for both designs, the 4 samples/cycle design increases the area cost by 46.5\% instead of doubling it and this is because the delay line is reduced. We notice that the percentage of area increase for the \gls{lfnst} is lower than for the \gls{mts}. This caused by the delay line of the \gls{lfnst} which is larger than in the \gls{mts}.

\subsubsection{VVC Transform vs HEVC Transform}

To compare the \gls{vvc} transform and the \gls{hevc} transform, we synthesised two 1-D design that are similar in throughput. The 1-D \gls{hevc} transform supports all \gls{dct}-II and \gls{dst} for block size $4 \plh 4$, it has a throughput of 2 samples/cycle with a static system latency. The \gls{vvc} transform supports all \gls{mts} types and sizes and all \gls{lfnst} sizes with a throughput of 2 samples/cycle and a fixed system latency. The difference between the two transforms is that the \gls{vvc} one is designed based on a \gls{rm} architecture while \gls{hevc} transform is designed using a \gls{mcm} based architecture. Many study proved that using \gls{mcm} based architecture for \gls{hevc} transform is more efficient in terms of area than using \gls{rm} based one. However, based on our previous study in \cite{9054281}, it turns out that for \gls{vvc} using \gls{rm} based architecture is more area efficient due to the new transform types and sizes. This fact proves that both architectures are optimally designed and thus giving more credibility for fair comparison.     
Table \ref{tab:synVVC-HEVC} shows the performance results of both designs. We can notice that the area of \gls{vvc} transform is 4 time larger than the \gls{hevc} transform block, this is considered to be reasonable rather good due of the new tools and new sizes introduced in \gls{vvc}.

\subsubsection{Hardware synthesis performance}

The proposed design supports the transform block of three different video standards including AVC/H.264, HEVC/H.265 and the emerging VVC/H.266 standard. The 1-D \gls{mts} core operates at 600 MHz with a total of 89K cell area and 17408-bits of \gls{rom} used to store \gls{mts} kernels. The \gls{lfnst} core operates at 600 MHz with a total of 74.5K cell area and 32768-bits of \gls{rom} used to store its kernels. Integrating these modules together adds two memories one for the transform coefficients (input) and the other for transformed samples (output). The input and output memories are \glspl{sram} that can hold a \gls{ctu} of size $64\plh64$ for 4:2:2 chroma subsampling. The Input memory is used as transpose memory for the 2-D \gls{mts}. It stores one Luma \gls{ctb} of size $64\plh64$ and two Chroma \glspl{ctb} of size $64\plh32$. The input sample size is 18-bits which makes the total input memory size 147456-bits. The output sample size is 11-bits resulting in a total of 90112-bits for the output memory. 

A fair comparison with the state of the art solutions is quite difficult since most of works focus on earlier versions of the \gls{vvc} \gls{mts} and do not support the \gls{lfnst}. \adcomment{Table \ref{CompareMTS} and \ref{comparison2D} give the key performance of state-of-the-art \gls{fpga} and \gls{asic}-based works, respectively. The only work that supports all \gls{mts} types and sizes is found in \cite{9081944} and its performance is presented in Table \ref{CompareMTS} for \gls{fpga} platform. Compared to our \gls{mts} design on \gls{fpga} platform, we consume more resources in terms of \glspl{alm} and registers, but on the other hand, we do not use any additional memory. Both designs support the complete version of \gls{vvc} \gls{mts} with similar performance, however, in this paper we add the \gls{lfnst} tool which requires nearly the same hardware resource as the \gls{mts} module.

The solutions proposed in~\cite{6571937, Can-mert, 8794833} support only the \gls{mts} module on \gls{asic} platform. Gate count is the logical calculation part and it can be seen from Table~\ref{comparison2D} that compared with implementations of Jia {\it et al.}~\cite{6571937}, Mert~{\it et al.} \cite{Can-mert} and Fan {\it et al.}~\cite{8794833}, our solution has several advantages. We present a unified transform architecture that can process inverse \gls{mts} and \gls{lfnst} supporting all kernel sizes including asymmetric sizes and all types (\gls{idct}-II/VIII and \gls{idst}-VII) in the two dimensions. Although our design supports more transform types and sizes than~\cite{6571937}, we still reduce the total area by up to 44\% and that's due to our multiplications design relying on shared regular multipliers. Compared to \cite{Can-mert} and \cite{8794833}, our design requires 3 times less area.}

\section{Conclusion}
\label{sec:conclusion}
In this paper a hardware implementation of the inverse \gls{vvc} transform module has been investigated targeting a real time \gls{vvc} decoder on \gls{asic} platforms. The \gls{vvc} transform module consists of a primary transform block and a secondary transform block called \gls{mts} and \gls{lfnst}, respectively. The proposed hardware implementation relies on regular multipliers and sustains a constant system latency with a fixed throughput of 1 cycle per sample. The \gls{mts} design leverage all primary transform optimisations including butterfly decomposition for \gls{dct}-II, zeroing for 64-point \gls{dct}-2 and 32-point \gls{dst}-VII/\gls{dct}-VIII, and the linear relation between \gls{dst}-VII and \gls{dct}-VIII. The proposed design uses 64 regular multipliers (32 for \gls{mts} and 32 for \gls{lfnst}) enabling to reach real time decoding of 4K video (4:2:2) at 30 frames par second. This architecture is scalabale and can easily be extended to reach a higher frame rate of 60 frames per second by using 128 regular multipliers. 

The proposed transform design has been successfully integrated in a hardware \gls{asic} decoder supporting the transform module of recent MPEG standards including AVC, HEVC and VVC. The hardware \gls{asic} decoder will be integrated in consumer devices to decode AVC, HEVC and VVC videos.








%
\bibliographystyle{IEEEtran}
\bibliography{IEEEexample}

%

\begin{IEEEbiography}[{\includegraphics[width=1in,height=1.25in,clip,keepaspectratio]{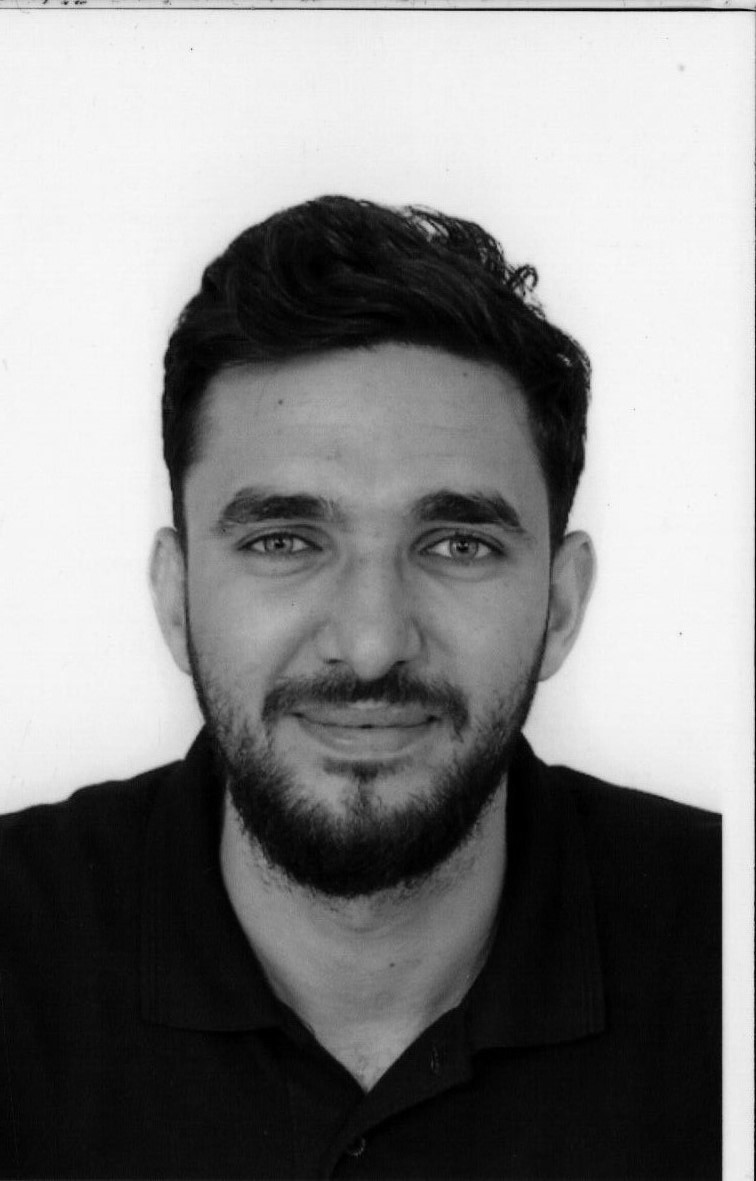}}]{Ibrahim Farhat}
was born in Eljem, Tunisia, in 1993. He received the engineering degree in communication systems and computer science from SUP'COM school of engineering, Tunis, in 2018. In 2019, he joined the Institute of Electronic and Telecommunication of Rennes (IETR), Rennes, and became a member of VITEC company hardware team, France, where he is currently pursuing the Ph.D. degree. His research interests focus on video coding, efficient real time and parallel architectures for the new generation video coding standards, and ASIC/FPGA hardware implementations.    
\end{IEEEbiography}
\vspace{-5mm}
\begin{IEEEbiography}[{\includegraphics[width=1in,height=1.25in,clip,keepaspectratio]{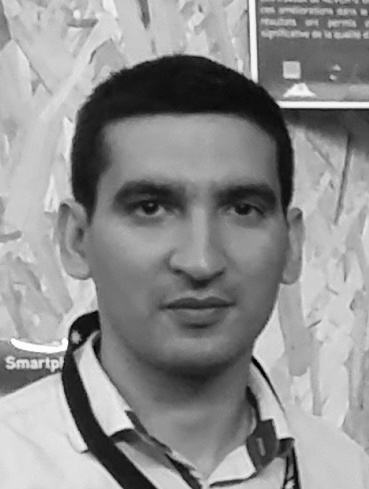}}]{Wassim Hamidouche}
received Master’s and Ph.D. degrees both in Image Processing from the University of Poitiers (France) in 2007 and 2010, respectively. From 2011 to 2013, he was a junior scientist in the video coding team of Canon Research Center in Rennes (France). He was a post-doctoral researcher from Apr. 2013 to Aug. 2015 with VAADER team of IETR where he worked under collaborative project on HEVC video standardisation. Since Sept. 2015 he is  an Associate Professor at INSA Rennes and a member of the VAADER team of IETR Lab. He has joined the Advanced Media Content Lab of b$<>$com IRT Research Institute as an academic member in Sept. 2017. His research interests focus on video coding and multimedia security. He is the author/coauthor of more than one hundred and forty papers at journals and conferences in image processing, two MPEG standards, three patents, several MPEG contributions, public datasets and open source software projects.
\end{IEEEbiography}
\vspace{-10mm}
\begin{IEEEbiography}[{\includegraphics[width=1in,height=1.25in,clip,keepaspectratio]{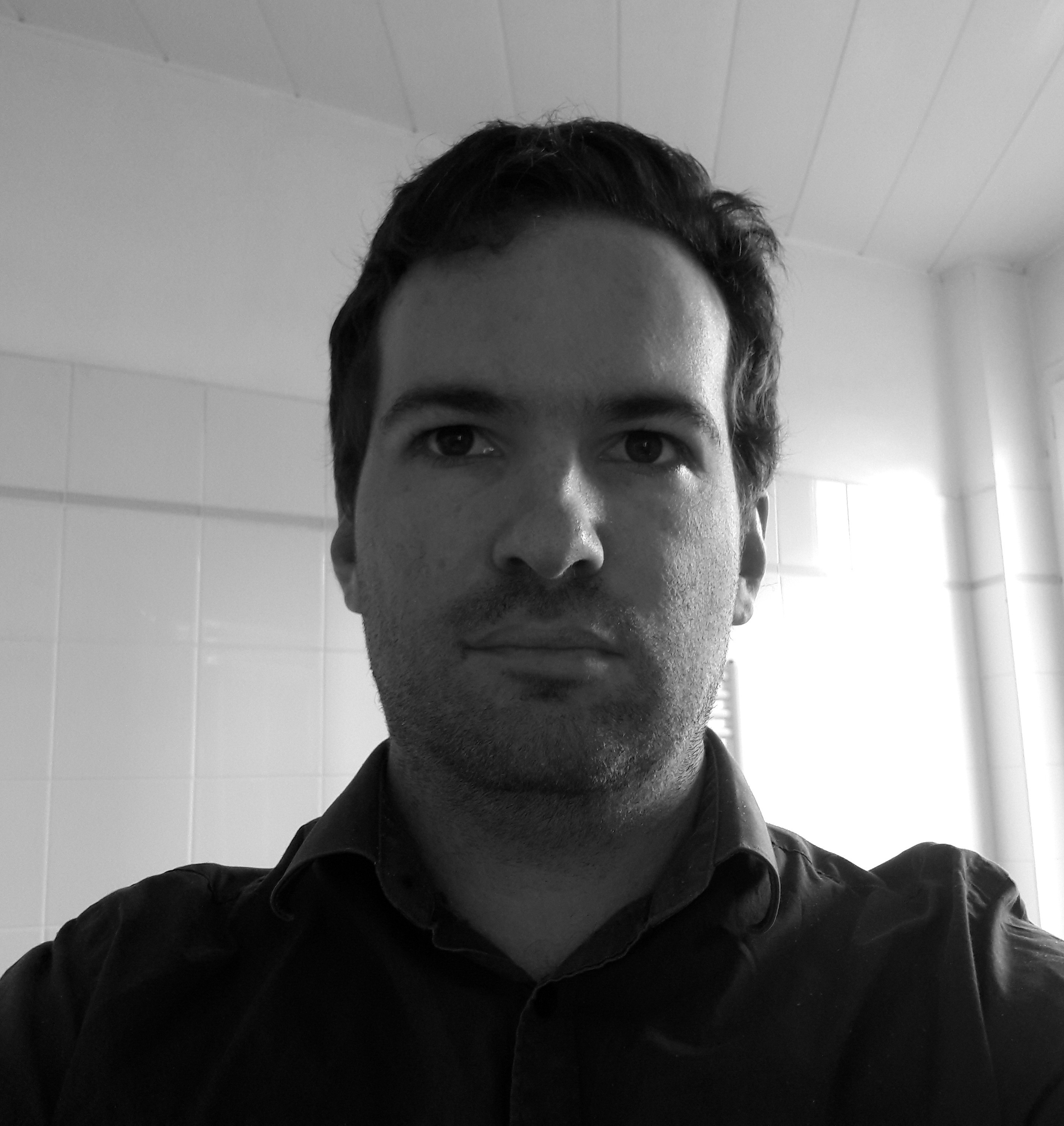}}]{Adrien Grill}
was born in 1988 in Aix-en-Provence, France. After receiving his engineering degree at Supelec in 2010, he specialized in hardware algorithm implementation. He joined Vitec in 2014, and currently works as technical leader on codec implementation projects. His fields of interest include signal, image processing, and video coding.
\end{IEEEbiography}
\vspace{-10mm}
\begin{IEEEbiography}[{\includegraphics[width=1in,height=1.25in,clip,keepaspectratio]{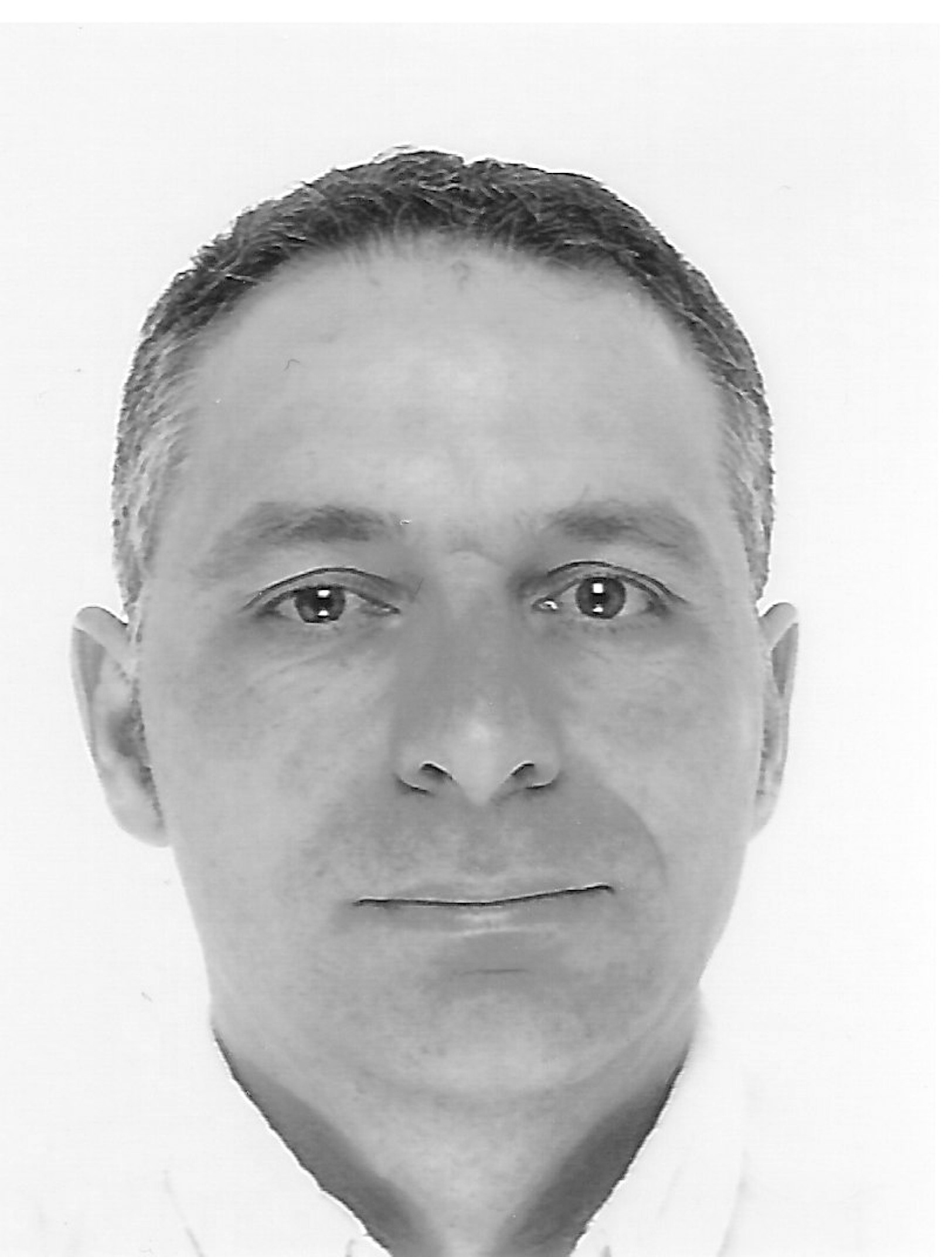}}]{Daniel Ménard}
received the Ph.D. and Habilitation degrees in Signal Processing and Telecommunications from the University of Rennes, respectively in 2002 and 2011. Since 2012, he has been Full-Professor at INSA Rennes - department of Electrical and Computer Engineering and member of the IETR lab. He has 20 years of expertise in the design and implementation of image and signal processing systems. His research interests include low power video codecs, approximate computing and energy consumption.
He has a long experience of collaborative projects, he has been involved in different national and European projects He is currently member of different Technical Program Committees of international conferences (ICASSP, SiPS and DATE). Since 2018, he has been an elected member of the Technical Committee ASPS of the IEEE Signal Processing society. He has published more than 100 scientific papers in international journal and conferences.
\end{IEEEbiography}
\vspace{-5mm}
\begin{IEEEbiography}[{\includegraphics[width=1in,height=1.25in,clip,keepaspectratio]{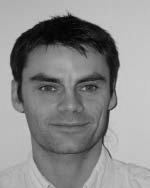}}]{Olivier D\'eforges}
received the Ph.D. degree in image processing in 1995. He is a Professor with the National Institute of Applied Sciences (INSA) of Rennes. In 1996, he joined the Department of Electronic Engineering, INSA of Rennes, Scientic and Technical University. He is a member of the Institute of Electronics and Telecommunications of Rennes (IETR), UMR CNRS 6164 and leads the IMAGE Team, IETR Laboratory including 40 researchers. He has authored over 130 technical papers. His principal research interests are image and video lossy and lossless compression, image understanding, fast prototyping, and parallel architectures. He has also been involved in the ISO/MPEG standardization group since 2007.
\end{IEEEbiography}


\end{document}